\documentclass{article}

\usepackage{PRIMEarxiv}

\usepackage[utf8]{inputenc} 
\usepackage[T1]{fontenc}    
\usepackage{hyperref}       
\usepackage{url}            
\usepackage{booktabs}       
\usepackage{amsfonts}       
\usepackage{nicefrac}       
\usepackage{microtype}      
\usepackage{lipsum}
\usepackage{fancyhdr}       
\usepackage{graphicx}       
\graphicspath{{media/}}     
\usepackage{subcaption}
\usepackage{amssymb}
\usepackage{amsmath}

\title{Learning to Predict RNA Sequence Expressions\\ from Whole Slide Images\\ with Applications for Search and Classification
}

\author{Amir Safarpoor$^1$, Jason D. Hipp$^2$, H.R. Tizhoosh$^{1,3,}$\thanks{Corresponding author} \\
  $^1$Kimia Lab, University of Waterloo, 
  Waterloo, ON, Canada\\
  $^2$   Division of Computational Pathology and AI, Mayo Clinic, Rochester, MN, USA\\
  $^3$   Department of Artificial Intelligence and Informatics, Mayo Clinic,  Rochester, MN, USA \\
  \texttt{asafarpoor@uwaterloo.ca, hipp.jason@mayo.edu, tizhoosh.hamid@mayo.edu}}

\begin{document}
\maketitle

\begin{abstract}
  Deep learning methods are widely applied in digital pathology to address clinical challenges such as prognosis and diagnosis. As one of the most recent applications, deep models have also been used to extract molecular features from whole slide images. Although molecular tests carry rich information, they are often expensive, time-consuming, and require additional tissue to sample. In this paper, we propose \textbf{tRNAsfomer}, an attention-based topology that can learn both to predict the bulk RNA-seq from an image and represent the whole slide image of a glass slide simultaneously. The \textbf{tRNAsfomer} uses multiple instance learning to solve a weakly supervised problem while the pixel-level annotation is not available for an image. We conducted several experiments and achieved better performance and faster convergence in comparison to the state-of-the-art algorithms. The proposed \textbf{tRNAsfomer} can assist as a computational pathology tool to facilitate a new generation of search and classification methods by combining the tissue morphology and the molecular fingerprint of the biopsy samples.  
\end{abstract}


\section{Introduction}

Pathologists use histopathology to diagnose and grade cancer after examining a biopsy specimen. The introduction of digital pathology, advances in computing technology, and the expanding availability of massive datasets made it possible to train increasingly complex deep learning models for various clinical tasks. Convolutional neural networks (CNNs) surpassed all other traditional computer vision algorithms in a wide range of clinical applications, including cancer subtyping \cite{hou2016patch}, whole-slide image (WSI) search and categorization \cite{kalra2020pan}, mitosis detection \cite{wang2014mitosis}, and grading \cite{bulten2020automated}, among deep learning architectures.

However, there have been a few attempts to connect the morphological characteristics embedded in the images to molecular signatures, recently \cite{schmauch2020deep,levy2020spatial,he2020integrating,tavolara2021deep}. For instance, recent research has revealed that statistical models can link histomorphological traits to mutations in organs, including the lung and prostate \cite{coudray2018classification,schaumberg2017h}. Mutations and epigenomic modifications are known to cause large variations in gene expression. Therefore, characterization of the gene expression can be vital for diagnosis and treatment \cite{segal2005signatures}. Even though more affordable whole transcriptome sequencing tools for studying gene information have been established, they are still a long way from being widely used in medical centers \cite{kamps2017next}. On the other hand, the recovery of molecular features from hematoxylin \& eosin (H\&E) stained WSIs is one of the faster and less expensive options. The capability to predict gene expression using WSIs, either as an intermediate modality or as an outcome, has been demonstrated to aid diagnosis and prognosis \cite{schmauch2020deep,tavolara2021deep}. Previous studies have drawn attention to gene expression prediction using WSI; however, the size of WSIs and the amount of well-annotated data still impose serious challenges. In particular, sample selection and WSI representation is an open topic that is often handled arbitrarily.

According to the most recent global cancer statistics report, in $2020$, there were an estimated $431,288$ new cases of kidney cancer and $179,368$ deaths globally \cite{sung2021global}. The RCC is the most common kidney cancer that is responsible for $85\%$ malignant cases \cite{shuch2015understanding}. From a single malignant phenotype to a heterogeneous group of tumors, our knowledge about RCC has evolved over time \cite{shuch2015understanding}. Among all RCC histologic subtypes, ccRCC, pRCC, and crRCC make almost $75\%$, $16\%$, and $7\%$ of the whole RCC cases, respectively \cite{shuch2015understanding}. RCC subtypes differ in their histology, molecular characteristics, clinical outcomes, and therapeutic responsiveness as a result of this heterogeneity. For instance, because the 5-year survival rate differs across different subtypes, proper subtype diagnosis is critical \cite{tabibu2019pan}. All methods in this work are applied on RCC slides to identify the subtypes using search and classification.

Here, we introduce \textbf{tRNAsformer} (pronounced \emph{t-RNAs-former}), a deep learning model for end-to-end gene prediction and learning WSI representation at the same time. Our model employs transformer modules built on the attention mechanism to gather information required for learning a WSI representation. To train our model, we used data from \emph{The Cancer Genome Atlas} (TCGA) public dataset to gather kidney WSIs and their related RNA-seq data. For WSIs, we presented our findings related to gene prediction and internal representation. Finally, we tested the generalization of our model in terms of learned WSI internal representation against state-of-the-art benchmarks using an external kidney cancer dataset from the Ohio State University.

\section{Results}

\paragraph{A model for predicting gene expression from WSIs.} The FPKM-UQ files containing 60,483 Ensembl gene IDs were utilized in this study \cite{hubbard2002ensembl}. During the preprocessing step (described in \ref{sec:gene-preprocessing}), some of the gene expression values were selected and then transformed first.

\begin{figure}
     \centering
     \begin{subfigure}[b]{1\textwidth}
         \centering
         \includegraphics[width=0.8\textwidth]{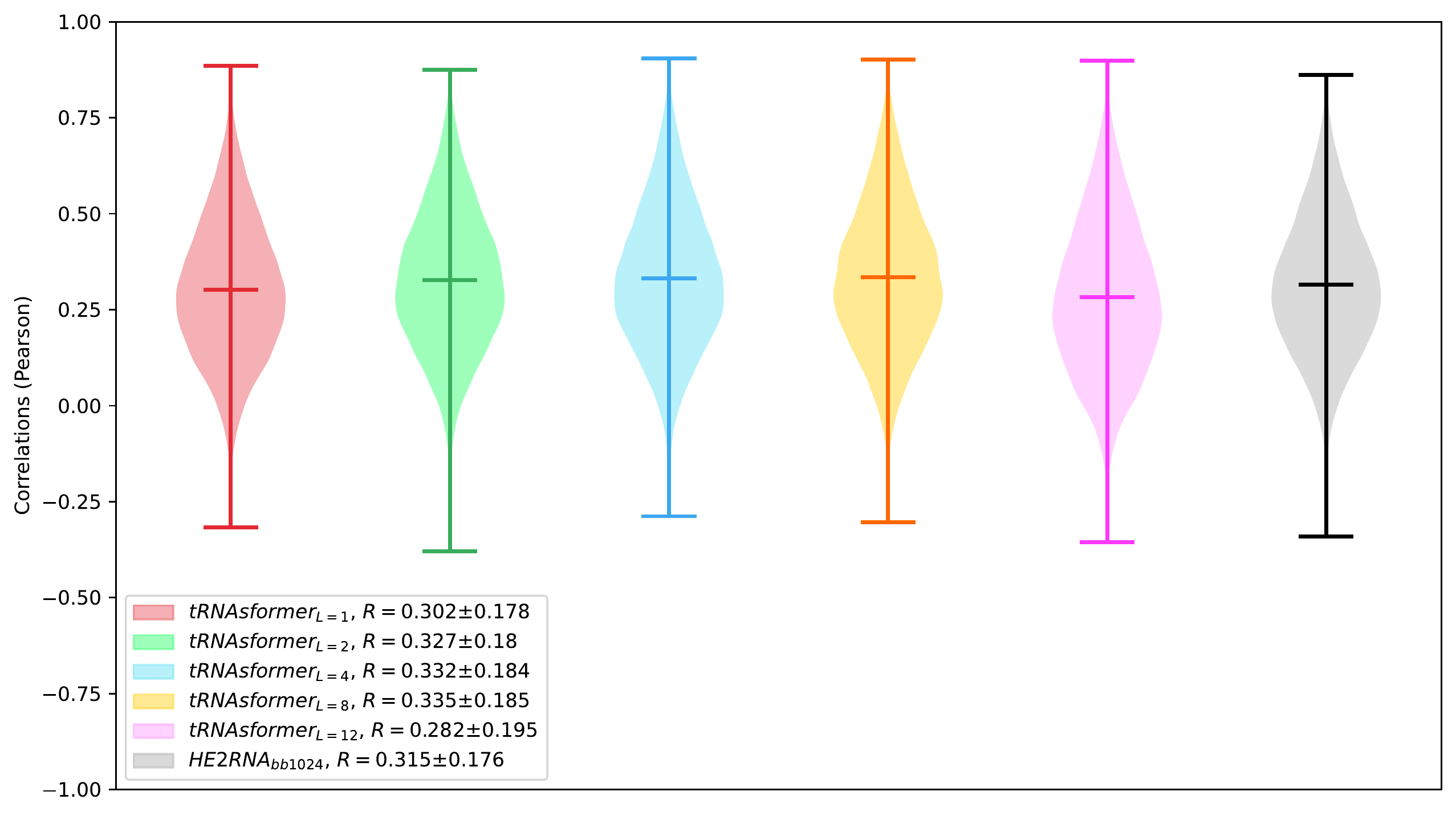}
         \caption{Pearson correlation.}
         \label{fig:rna-pearson}
     \end{subfigure}\\
     \begin{subfigure}[b]{1\textwidth}
         \centering
         \includegraphics[width=0.8\textwidth]{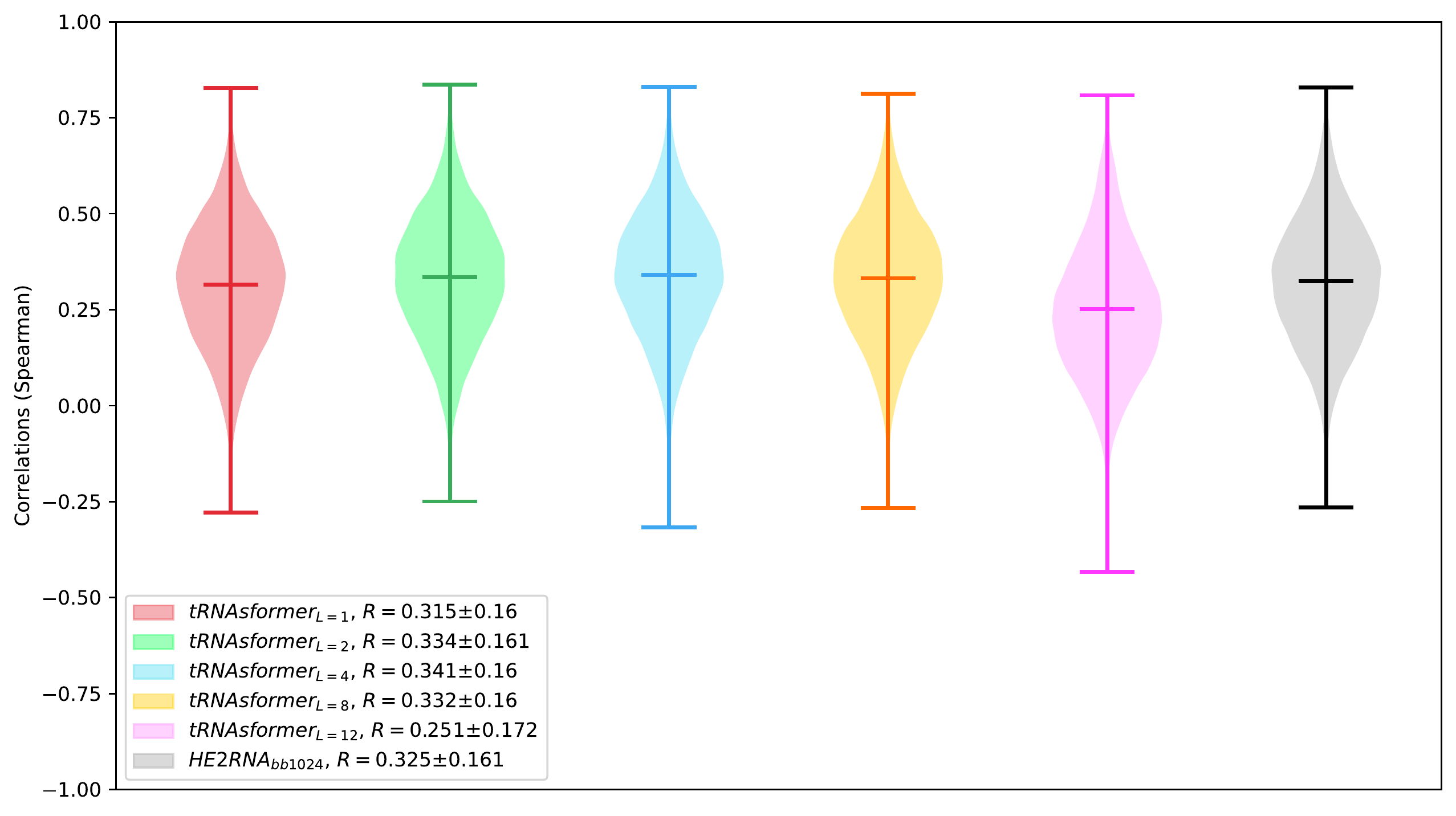}
         \caption{Spearman correlation.}
         \label{fig:rna-spearman}
     \end{subfigure}
        \caption{The distribution of the correlation coefficients between 31,793 genes predicted and their true value for TCGA test set. The violin diagrams depict the distribution, min, max, and mean values of the correlation coefficients. (a) violin diagrams for Pearson correlation coefficients and (b) violin diagrams for Spearman's correlation coefficients. The violin diagrams are plotted for $\textrm{tRNAsformer}_{L}$ for $L=(1,2,4,8,12)$ and $\textrm{HE2RNA}_{\textrm{bb}1024}$. The mean and standard deviation of the correlation coefficients are included in the legend.}
        \label{fig:rna-correlation}
\end{figure}

Both models, \textbf{tRNAsformer} and HE2RNA, were compared for three different criteria, namely mean correlation coefficient of predictions, the number of genes predicted significantly better than a random baseline, and the prediction error. In the first experiment, the correlation is assessed for each gene separately using Pearson and Spearman's correlation coefficient. If the datasets are normally distributed, the Pearson correlation coefficient measures the linear connection between them. The Pearson correlation coefficient varies between $-1$ and $+1$. A correlation of $-1$ or $+1$ denotes a perfect linear negative or positive relationship, respectively, whereas a correlation of 0 denotes no correlation. The $p$-value roughly represents the probability that an uncorrelated system can produce datasets with a Pearson correlation at least as high as the one calculated from these datasets. The Spearman correlation, unlike the Pearson correlation, does not require that both datasets be normally distributed. Fig. \ref{fig:rna-correlation} displays the distribution of correlation coefficient for 31,793 genes predicted by different models.

The mean correlation coefficient $R$ grew with depth from $L=1$ to $L=8$, as seen in Fig. \ref{fig:rna-correlation}. The mean $R$ value declines after eight blocks of Transformer encoders, suggesting that increasing the number of layers does not enhance gene expression predictions. Another important observation is that \textbf{tRNAsformer} has higher mean correlation coefficients than its counterpart for $L=2$ to $L=8$.

The Pearson and Spearman's correlation coefficients and $p$-values were computed between the predicted and the true value of the gene expression for each gene. Two multiple-hypothesis testing methods, namely Holm-Šidák (HS) and Benjamini-Hochberg (BH), were utilized to adjust the $p$-values. If the $p$-value of the $R$ coefficient was less than $0.01$ after correction for multiple-hypothesis testing, the prediction was significantly different from the random baseline \cite{holm1979simple,benjamini1995controlling}. Similar to \cite{schmauch2020deep}, multiple-hypothesis testing was done using both HS and BH correction. The results are shown in Table \ref{tab:rna-multiple-test} for all architectures.

\begin{table}
    \centering
    \caption{The number of genes were predicted with a statistically significant correlation ($p$-value $< 0.01$) under HS and BH correction. The total number of predicted genes is 31,793. These values are computed using the TCGA test dataset.} \label{tab:rna-multiple-test}
    \begin{tabular}{lcccc}
    \hline
    & \multicolumn{2}{c}{Pearson} & \multicolumn{2}{c}{Spearman} \\ 
    \cmidrule(lr){2-3} \cmidrule(lr){4-5}
    Model   & HS     & BH    & HS     & BH     \\ \hline
    $\textrm{tRNAsformer}_{L=1}$    & 29,990          & 30,797      & 30,427          & 31,042  \\
    $\textrm{tRNAsformer}_{L=2}$    & 30,338          & 31,014      & 30,695          & 31,141  \\
    $\textrm{tRNAsformer}_{L=4}$    & 30,433          & 30,996      & 30,858          & 31,266  \\ 
    $\textrm{tRNAsformer}_{L=8}$    & 30,344          & 31,002      & 30,741          & 31,181  \\
    $\textrm{tRNAsformer}_{L=12}$   & 28,933          & 30,187      & 28,938          & 30,210  \\
    \hline
    $\textrm{HE2RNA}_{\textrm{bb}1024}$ &   30,249    & 30,937  &   30,663  &   31,163  \\
    \hline
\end{tabular}
\end{table}

As it is demonstrated in Table \ref{tab:rna-multiple-test}, increasing the depth of the \textbf{tRNAsformer} from one to eight increases the number of genes that are significantly different from a random baseline. Similar to the results in Fig. \ref{fig:rna-correlation}, there is a decrease in the number of genes when the depth reaches 12 blocks of Transformer Encoder. On the other hand, the model based on the design of $\textrm{HE2RNA}$ scored inferior to nearly all other \textbf{tRNAsformer} models (except for $L=1$).

We selected MAE, RMSE, and RRMSE \cite{spyromitros2016multi} to calculate the error between the prediction and real gene expression values. MAE, RMSE, and RRMSE are defined as

\begin{align}
    \textrm{MAE} &= \frac{ \sum_{(x_i, y_i) \in D_{\textrm{test}}} |\hat{y}_i - y_i|}{|D_{\textrm{test}}|},\\
    \textrm{RMSE} &= \sqrt{ \frac{ \sum_{(x_i, y_i) \in D_{\textrm{test}}} (\hat{y}_i - y_i)^2}{|D_{\textrm{test}}|} },\\
    \textrm{RRMSE} &= \sqrt{ \frac{ \sum_{(x_i, y_i) \in D_{\textrm{test}}} (\hat{y}_i - y_i)^2 }{ \sum_{(x_i, y_i) \in D_{\textrm{test}}} (\bar{y} - y_i)^2 } },
\end{align}

\noindent where $D_{\textrm{test}}$ denotes the test set, $(x_i, y_i)$ is the $i$-th sample $x_i$ with ground truth $y_i$, $\hat{y}_i$ is the predicted value of $y_i$, $\bar{y}$ is the mean value over the targets in the test set, and $|D_{\textrm{test}}|$ is the number of samples in the test set. The results are given in Table \ref{tab:rna-error}.

\begin{table}
    \centering
    \caption{Prediction error for \textbf{tRNAsformer} and $\textrm{HE2RNA}_{\textrm{bb}1024}$ models quantified by MAE, RMSE, and RRMSE. All errors are calculated using TCGA test set.}\label{tab:rna-error}
    \begin{tabular}{lccc}
    \hline
    Model & MAE       & RMSE      & RRMSE     \\
    \hline
    $\textrm{tRNAsformer}_{L=1}$  & $1.31 \pm 1.04$ & $1.67 \pm 1.20$ & $1.02 \pm 0.16$ \\
    $\textrm{tRNAsformer}_{L=2}$  & $1.30 \pm 1.03$ & $1.65 \pm 1.17$ & $1.02 \pm 0.16$ \\
    $\textrm{tRNAsformer}_{L=4}$  & $1.30 \pm 1.08$ & $1.63 \pm 1.19$ & $0.98 \pm 0.11$  \\
    $\textrm{tRNAsformer}_{L=8}$  & $1.37 \pm 1.02$ & $1.69 \pm 1.13$ & $1.11 \pm 0.27$  \\
    $\textrm{tRNAsformer}_{L=12}$  & $1.50 \pm 1.10$ & $1.79 \pm 1.26$ & $1.10 \pm 0.43$  \\
    \hline
    $\textrm{HE2RNA}_{\textrm{bb}1024}$   &   $1.29 \pm 1.08$ & $1.63 \pm 1.20$ & $0.96 \pm 0.08$  \\
    \hline
    \end{tabular}
\end{table}

Similar to the results in Fig. \ref{fig:rna-correlation} and Table \ref{tab:rna-multiple-test}, increasing the number of Transformer Encoder blocks from eight to 12 significantly degrades the performance of the model. In addition, the \textbf{tRNAsformer} has a comparable for $L=1$ to $L=8$, considering the fact that \textbf{tRNAsformer} handles multiple tasks rather than a single gene prediction task. Overall, according to the correlations coefficients, p-value tests, and prediction errors, $L=8$ appears to be a critical threshold after which the model becomes overparametrized for the gene prediction task, according to the results.

\paragraph{Transcriptomic learning for WSI representation -- WSI classification.} The classification experiments were conducted to assess the quality of internal representation learned by the proposed model. To begin, 100 bags have been created from each TCGA test WSIs. According to Table \ref{tab:tcga-kidney}, a total of 8,000 bags were created from TCGA test set, as there were 80 WSIs. The same models that were trained in the previous section to predict RCC subtypes were assessed for the classification task as well. The accuracy, macro, and weighted F1 scores is presented for all models in Table \ref{tab:rna-acc-f1}. The confusion matrices of different models are displayed in Fig. \ref{fig:rna-confusion-tcga}. All values reported here are based on slide-level classification results. The prediction is made for all bags in order to calculate slide-level values. Each test slide's label predication is chosen as the most common prediction among all bags created from that slide. The WSI representations learned by the models are projected onto a plane created by the first two principal components found using PCA to depict the internal representation of our models in two-dimensional space. The two-dimensional PCA projections are shown in Fig. \ref{fig:rna-pca-tcga}.

Because of variations in hospital standards and methods for tissue processing, slide preparation, and digitization protocols, the appearance of WSIs might vary significantly. As a result, it is important to ensure that models built using data sources are resistant to data-source-specific biases and generalize to real-world clinical data from sources not used during training \cite{stacke2019closer}. For testing the generalization of our trained models, 142 RCC WSIs are used from the Ohio State University as an independent test cohort (see Section \ref{sec:external-dataset}).

First, 100 bags were created from each external test WSIs. According to Table \ref{tab:tcga-kidney}, a total of 14,200 bags were created from TCGA test set, as there were 142 WSIs. Same models that were trained in the previous section to predict RCC subtypes are used to report classification results for the external dataset. The accuracy, macro, and weighted F1 scores are reported for all models in Table \ref{tab:rna-acc-f1}. The confusion matrices of different models are displayed in Fig. \ref{fig:rna-confusion-external}. The WSI representations learned by the models are projected onto a plane created by the first two principal components found using PCA to depict the internal representation of the models in two-dimensional space. The two-dimensional PCA projections are shown in Fig. \ref{fig:rna-pca-external}.

\begin{table}
    \centering
    \caption{The accuracy, macro, and weighted F1 scores for classification on TCGA test set and the external dataset for all classification models.}\label{tab:rna-acc-f1}
    \begin{tabular}{lcccccccc}
        \hline
         & \multicolumn{3}{c}{TCGA}                     & \multicolumn{3}{c}{External dataset}                 \\
         \cmidrule(lr){2-4} \cmidrule(lr){5-7}
         &  & \multicolumn{2}{c}{F1 score} &  & \multicolumn{2}{c}{F1 score} \\
         \cmidrule(lr){3-4} \cmidrule(lr){6-7}
        Model & Accuracy & macro & weighted & Accuracy & macro & weighted \\
        \hline
        $\textrm{tRNAsformer}_{L=1}$ & 93.75\%   & 0.9488   & 0.9366      & 82.39\%   & 0.8241   & 0.8223     \\
        $\textrm{tRNAsformer}_{L=2}$ & 95.00\%   & 0.9406   & 0.9496      & 81.69\%   & 0.8161   & 0.8145     \\
        $\textrm{tRNAsformer}_{L=4}$ & 96.25\%   & 0.9511   & 0.9625      & 78.87\%   & 0.7899   & 0.7871     \\
        $\textrm{tRNAsformer}_{L=8}$ & 95.00\%   & 0.9414   & 0.9502      & 82.39\%   & 0.8251   & 0.8227     \\
        $\textrm{tRNAsformer}_{L=12}$ & 92.50\%   & 0.9392   & 0.9243      & 80.28\%   & 0.8072   & 0.8034    \\
        \hline
        Low power method \cite{safarpoor2021renal}    &   93.75\%   &   0.9488   &   0.9366   &   73.76\% &   0.7388  &   0.7385  \\
        \hline
    \end{tabular}
\end{table}

\begin{figure}
    \centering
    \begin{subfigure}{0.4\textwidth}
    \includegraphics[width=1\linewidth]{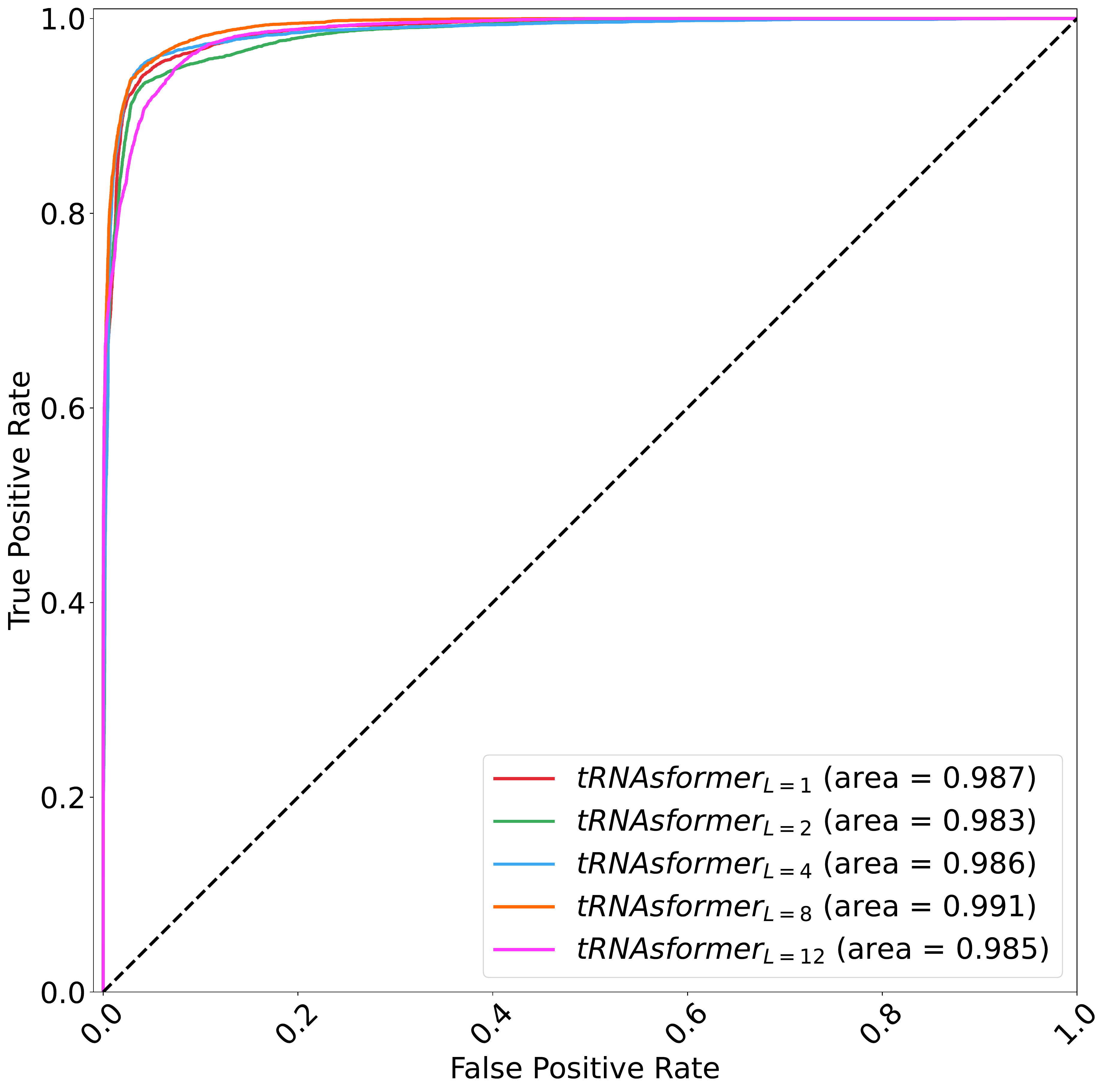}
    \caption{TCGA test dataset.}\label{fig:rna-roc-tcga}
    \end{subfigure}
    \begin{subfigure}{0.4\textwidth}
    \includegraphics[width=1\linewidth]{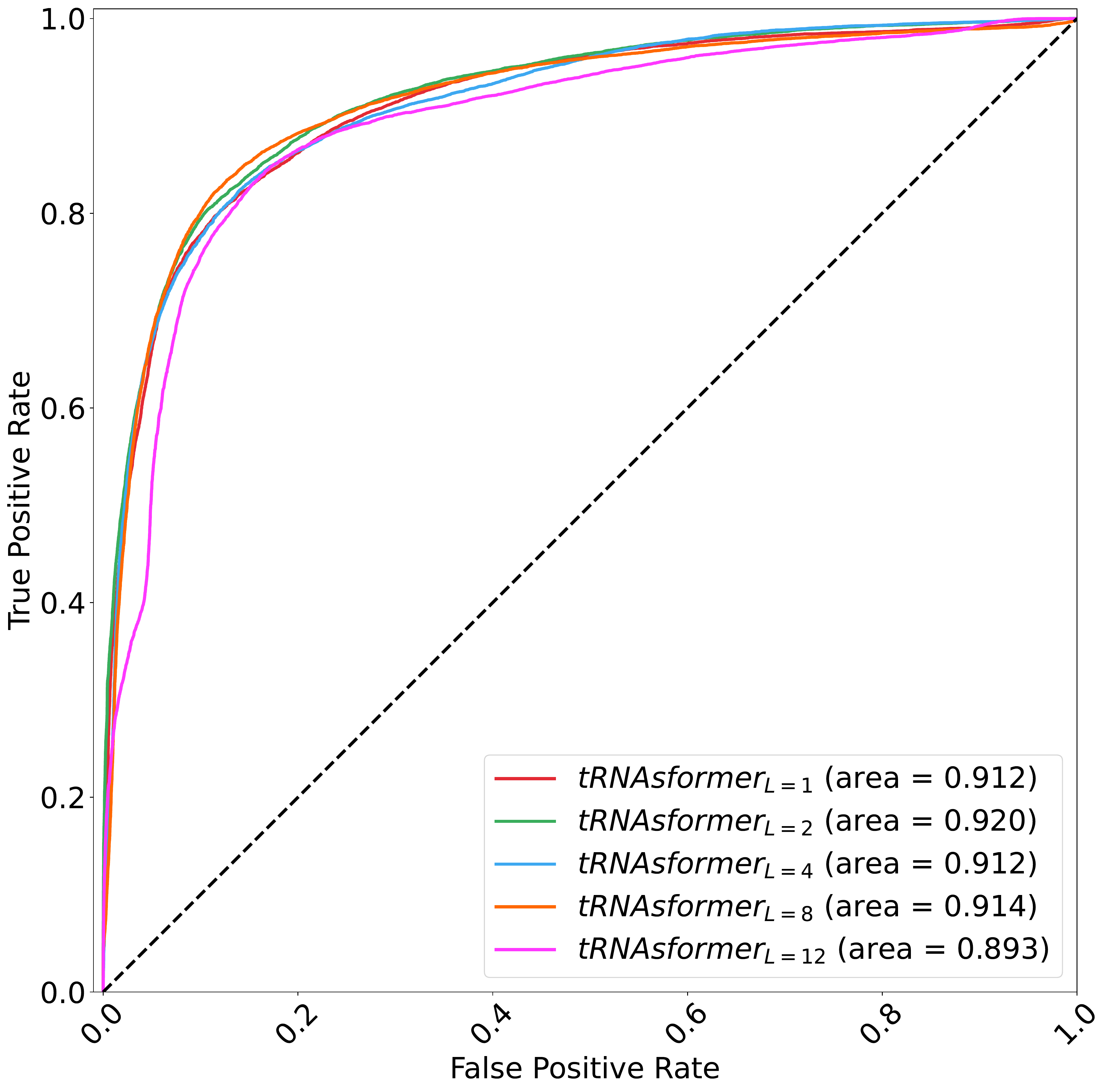}
    \caption{External dataset.}\label{fig:rna-roc-external}
    \end{subfigure}
    \caption{The micro ROC curve of different models applied on (a) TCGA test set and (b) the external dataset. The AUC is reported in the legend for all models.}
    \label{fig:rna-roc}
\end{figure}

The suggested model in \cite{safarpoor2021renal}, also known as the ``Low Power'' technique, outperformed all tile-based and state-of-the-art WSI-level approaches. The ``Low Power'' method's accuracy, F1 score (macro and weighted), and AUC were 73.76\%, 0.7388, 0.7385, and 0.893, respectively. As it is demonstrated in Table \ref{tab:rna-acc-f1} and Fig. \ref{fig:rna-roc}, all \textbf{tRNAsformer} models surpass the method described in \cite{safarpoor2021renal} in all measures, namely accuracy, F1 score (macro and weighted), and AUC. Additionally, as it is depicted in Fig. \ref{fig:rna-confusion-external}, the \textbf{tRNAsformer} models tend to have more balanced correct predictions for all classes as there is crisp diagonal line highlighted in confusion matrices. To put it another way, \textbf{tRNAsformer} models are good at distinguishing between all classes.

\paragraph{Transcriptomic learning for WSI representation -- WSI search.} WSI search experiments were conducted to assess the quality of the internal representation of the \textbf{tRNAsformer}. The model is tested on both TCGA and an external dataset. As it was mentioned earlier 100 instances were created from each WSI in TCGA dataset; TCGA test set contained 8,000 instances associated with 80 slides. To quantify the performance of \textbf{tRNAsformer} in WSI search, first, 100 subsets of instances were created from 8,000 TCGA test instances. Next, a pairwise distance matrix is computed using the WSI embeddings for each subset. The Pearson correlation is employed as the distance metric. Following the \emph{leave-one-patient-out} procedure, the top-$k$ samples were determined for each instance (WSI). Later, P@K and AP@K were computed for each subset. Finally, the MAP@K value was computed by taking average of 100 queries associated with 100 search subsets. 

Similarly, 100 instances were created for each WSI in the external dataset. Overall, 100 subsets of 142 WSIs generated for the WSI search in the external dataset. As a result, MAP@K values were evaluated by taking an average from 100 different search experiments. The summary of MAP@K values for both TCGA test and the external dataset are shown in Table \ref{tab:rna-search}.

\begin{table}
    \centering
    \caption{The MAP@5 and MAP@10 values for all WSI search models applied on TCGA test and the external dataset.}
    \label{tab:rna-search}
    \begin{tabular}{lcccc}
    \hline
    & \multicolumn{2}{c}{TCGA} & \multicolumn{2}{c}{External dataset} \\
    \cmidrule(lr){2-3}  \cmidrule(lr){4-5}
    Model                        & MAP@5                  & MAP@10                 & MAP@5             & MAP@10           \\ 
    \hline
    $\textrm{tRNAsformer}_{L=1}$ & 0.8966                 & 0.8985                 & 0.8026            & 0.8035           \\
    $\textrm{tRNAsformer}_{L=2}$ & 0.8831                 & 0.8800                 & 0.7988            & 0.7976           \\
    $\textrm{tRNAsformer}_{L=4}$ & 0.9150                 & 0.9124                 & 0.7819            & 0.7781           \\
    $\textrm{tRNAsformer}_{L=8}$ & 0.9031                 & 0.8996                 & 0.7674            & 0.7628           \\
    $\textrm{tRNAsformer}_{L=12}$ & 0.8762                 & 0.8751                 & 0.7262            & 0.7257           \\
    \hline
    Yottixel \cite{kalra2020pan}    &   0.764   &   0.717   &   0.7416  &   0.7092 \\
    \hline
    \end{tabular}
\end{table}

To compare \textbf{tRNAsformer}'s search results with Yottixel \cite{kalra2020yottixel}, the state-of-the-art in WSI search, the MAP@5 and MAP@10 for Yottixel were calculated. The MAP@5 and MAP@10 for 10 independent Yottixel runs were 0.7416 and 0.7092, respectively. \textbf{tRNAsformer} outperforms Yottixel in both MAP@5 and MAP@10 measures. Furthermore, \textbf{tRNAsformer} models provide more stability because the MAP@K value does not drop as steeply as other search algorithms while the $k$ increases.

\section{Discussion}

In this paper, a multitask MIL framework based on \textbf{tRNAsformer} model is proposed for learning WSI representation by learning to predict gene expression from H\&E slides. By incorporating the attention mechanism and the Transformer design, \textbf{tRNAsformer} can provide more precise predictions for gene expressions from a WSI. Meanwhile, \textbf{tRNAsformer} surpassed benchmarks for bulk RNA-seq prediction while having fewer hyperparameters. Additionally, \textbf{tRNAsformer} learns exclusive\footnote{A dedicated paradigm for distilling the bag information into a feature vector for WSI representation.} and compact representation for a WSI using molecular signature of the tissue sample. As a result, the proposed technique learns a diagnostically relevant representation from an image by integrating gene information in a multimodal approach.

Furthermore, the Transformer design allowed for more efficient and precise processing of a collection of samples. This property eliminates the need for costly and time-consuming pixel-by-pixel human annotations. Finally, sampling and embedding image tiles using pre-trained CNN models offers several advantages:

\begin{itemize}
    \item Trained on large image datasets, deep CNNs can be exploited to create rich intermediate embeddings from image samples.
    \item Working with embedded sampled instances\footnote{The tile deep features instead of tiles.} is computationally less expensive in comparison with treating each WSI as an instance. According to Table \ref{tab:rna-model-parameters}, the smallest \textbf{tRNAsformer} model can have about 60\% less hyperparameter in comparison with \textit{MLP}-based model. Additionally, they can be about 72\% and 15\% faster than \textit{MLP}-based model during training and validation, respectively.
    \item By augmenting data, bootstrapping meets the requirement for big datasets for training deep models.
    \item By diversifying the instances in a bag, bootstrapping at test time reduces noise.
\end{itemize}

In contrast to \cite{he2020integrating} where the spatial transcriptomics dataset was available, the proposed approach in this work uses bulk RNA-seq data. As a result, the model described in this study employs a weaker type of supervision, as it learns internal representation using a combination of a primary diagnosis and a bulk RNA-seq associated with a WSI. This is more in line with current clinical practice, which generally collects bulk RNA sequences rather than spatial transcriptomic data. Furthermore, \textbf{tRNAsformer} handles the problem by treating a WSI in its entirety, whereas the method explained in \cite{he2020integrating} separates each tile and estimates the gene expression value for it. Therefore, the method described in\cite{he2020integrating} ignores the dependencies between tiles. Comparing to \cite{tavolara2021deep}, the proposed technique in this manuscript processes a considerably smaller set of samples with a larger field of view. In particular, the proposed technique samples bags of 49 instances of $224 \times 224 \times 3$ while the other technique \cite{tavolara2021deep} deployed several sampling options with at least 2,500 tiles of size $32 \times 32 \times 3$ per bag. In addition, \textbf{tRNAsformer} learns exclusive WSI representation by learning the pixel-to-gene translation. On the other hand, none of the methodologies have an independent representation learning paradigm \cite{schmauch2020deep,he2020integrating,tavolara2021deep}.

In conclusion, the proposed framework can learn reliable internal representations for massive archives of pathology slides that match or outperform the performance of cutting-edge classification and search algorithms developed \cite{kalra2020yottixel,safarpoor2021renal}. It can also predict gene expressions from H\&E slides better than other methods \cite{schmauch2020deep}. By employing a balanced architecture, the proposed model outperforms existing topologies in both tasks simultaneously.

\section{Methods}

\paragraph{TCGA kidney dataset.} The data used in this study came from the TCGA (\href{https://portal.gdc.cancer.gov/}{https://portal.gdc.cancer.gov/}). We acquired kidney cases for which both WSI and RNA-seq was available. We selected H\&E-stained formalin-fixed, paraffin-embedded (FFPE) diagnostic slides. The retrieved cases included three subtypes, clear cell carcinoma, ICD-O $8310/3$, (ccRCC), chromophobe type - renal cell carcinoma, ICD-O $8317/3$, (crRCC), and papillary carcinoma, ICD-O $8260/3$, (pRCC). For transcriptomic data, we utilized Fragments Per Kilobase of transcript per Million mapped reads upper quartile (FPKM-UQ) files. The detailed information regarding the cases are included in Table \ref{tab:tcga-kidney}. The data was split case-wise into train ($\%80$), validation ($\%10$), test ($\%10$) sets, respectively. In other words, each patient only belonged to one of the sets.

\paragraph{Gene expression preprocessing.}\label{sec:gene-preprocessing} The FPKM-UQ files contained $60,483$ Ensembl gene IDs. We excluded genes with a median of zero across all kidney cases and we left with $31,793$ genes. Other studies have adopted the same strategy to improve the interpretability of the results \cite{schmauch2020deep}. We used $a \rightarrow \log_{10}(1+a)$ transform to convert the gene expressions since the order of gene expression values changes a lot and can impact mean squared error only in the case of highly expressed genes \cite{schmauch2020deep}. 

\paragraph{WSI preprocessing.} The size of the digitized glass slides may be $100,000 \times 100,000$ in pixels or even larger. As a result, processing an entire slide at once is not possible with present technology. These images are commonly divided into smaller, more manageable pieces known as \emph{tiles}. Furthermore, large WSI datasets are generally weakly labelled since pixel-level expert annotation is costly and labour-intensive. As a result, some of the tiles may not carry information that is relevant to the diagnostic label associated with the WSI. Consequently, MIL may be suitable for this scenario. Instead of receiving a collection of individually labelled examples, the learner receives a set of labelled bags, each comprising several instances in MIL. For making \emph{bags of instances}, the first step is to figure out where the tissue boundaries are. Using the algorithm described in \cite{safarpoor2021renal}, the tissue region was located at the thumbnail ($1.25 \times$ magnification) while the background and the marker pixels were removed. Tiles of size 14 by 14 pixels were processed using the $1.25 \times$ tissue mask to discard those with less than $50\%$ tissue. Note that 14 by 14 pixel tiles at $1.25 \times$ is equivalent to area of $224 \times 224$ pixels at $20 \times$ magnification.

\begin{figure}
     \centering
     \begin{subfigure}[b]{0.3\textwidth}
         \centering
         \includegraphics[width=\textwidth]{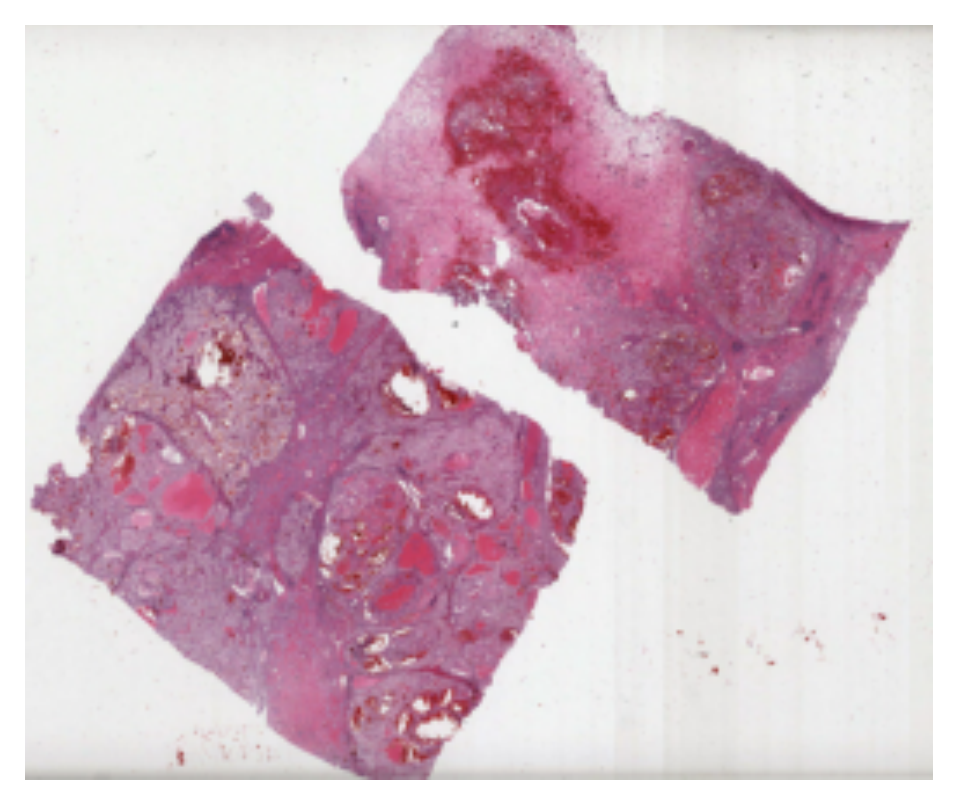}
         \caption{Thumbnail of a WSI.}
         \label{fig:thumbnail}
     \end{subfigure}
     \hfill
     \begin{subfigure}[b]{0.3\textwidth}
         \centering
         \includegraphics[width=\textwidth]{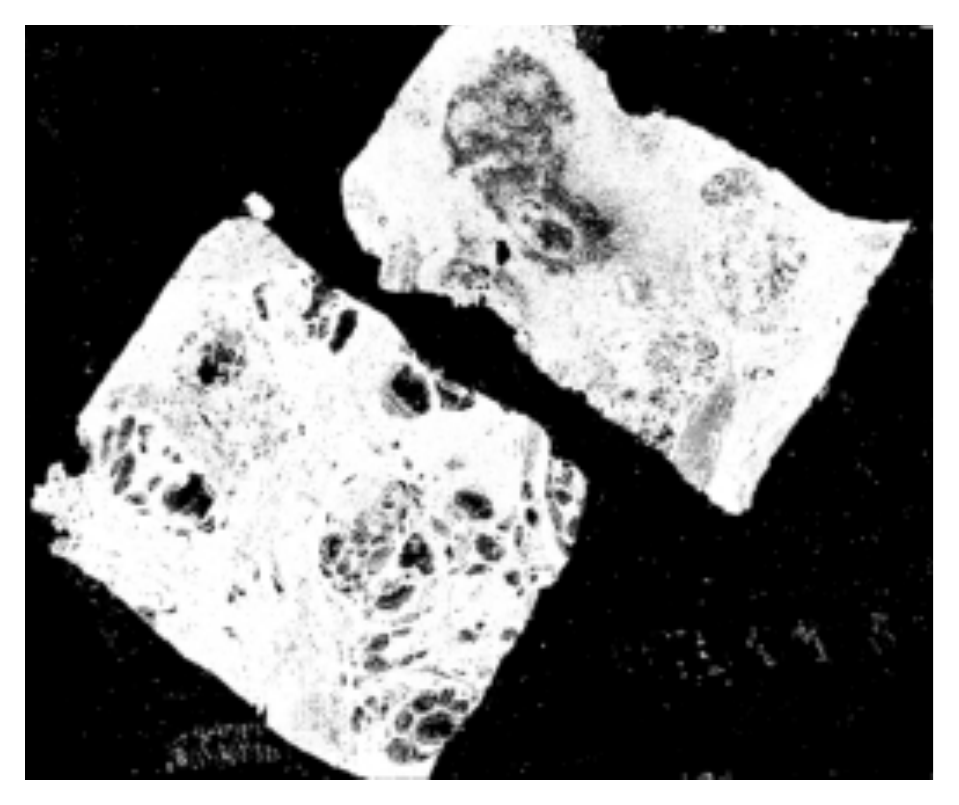}
         \caption{The tissue mask.}
         \label{fig:mask}
     \end{subfigure}
     \hfill
     \begin{subfigure}[b]{0.3\textwidth}
         \centering
         \includegraphics[width=\textwidth]{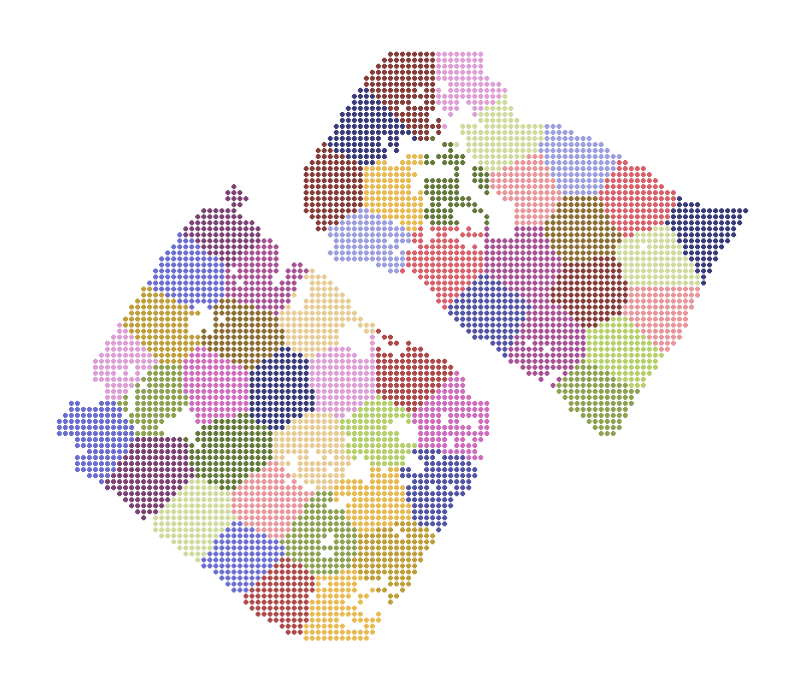}
         \caption{The $k$-means clusters.}
         \label{fig:clusters}
     \end{subfigure}
        \caption{An example of clustering for creating bag of tiles from a WSI.}
        \label{fig:clustering}
\end{figure}

The $k$-means algorithm is deployed on the location of the tiles selected previously to sample a fixed number of tiles from each WSI. The value of $k$ was set to 49 for all experiments in this study. After that, the clusters are spatially sorted based on the magnitude of the cluster centers. The benefit of spatially clustered tiles is twofold; (1) the concept of similarity is more likely to be true within a narrow radius \cite{sikaroudi2020supervision,gildenblat2019self}, and (2) clustering coordinates with two variables is computationally less expensive than high-dimensional feature vectors. The steps of the clustering algorithm are shown in Fig. \ref{fig:clustering}.

\paragraph{The tRNAsformer architecture.} The \textbf{tRNAsformer} is made of $L$ standard transformer encoder layers \cite{dosovitskiy2020image} followed by two heads, namely the classification and the gene prediction head. Fig. \ref{fig:diagram} depicts the architecture of the proposed method. The Transformer Encoder learns an embedding (also known as the \emph{class token}) for the input by treating it as a sequence of feature instances associated with each WSI. It learns internal embeddings for each instance while learning the class token that represents the bag or WSI.

The classification head, which is a linear layer, receives the WSI representation $\mathrm{\textbf{c}}$. Next, the WSI representation is projected using a linear layer to the WSI's score $\hat{y}$. \textbf{tRNAsformer} then uses cross-entropy loss between the predicted score $\hat{y}$ and the WSI's true label $\mathrm{\textbf{y}}$ to learn the primary diagnosis. The use of the Transformer Encoder and the classification head enables the learning of the WSI's representation while training the model.  

Considering a bag $\textrm{X} = [\mathrm{\textbf{x}}_{1}, \mathrm{\textbf{x}}_{2}, \dots, \mathrm{\textbf{x}}_{k}]$, where $\mathrm{\textbf{x}}_{i} \in \mathbb{R}^d$, $i=1,\dots,k$ are the embedded tiles by DenseNet-121, an $L$-layer standard Transformer can be defined as

\begin{align}
\mathrm{\textbf{z}}_{0} &= [\mathrm{\textbf{x}}_{class};\; \mathrm{\textbf{x}}_{1}\mathrm{\textbf{E}};\; \mathrm{\textbf{x}}_{2}\mathrm{\textbf{E}};\; \dots;\; \mathrm{\textbf{x}}_{k}\mathrm{\textbf{E}})] + \mathrm{\textbf{E}}_{pos},\;\;\;\;\;\;\;\;\;\; \mathrm{\textbf{E}} \in \mathbb{R}^{d \times D},\; \mathrm{\textbf{E}}_{pos} \in \mathbb{R}^{(k + 1) \times D}\\
\mathrm{\textbf{z}}^{\prime}_{\ell} &= \mathrm{MSA}(\mathrm{LN}(\mathrm{\textbf{z}}_{\ell-1})) + \mathrm{\textbf{z}}_{\ell-1},\;\;\;\;\;\;\;\;\;\;\;\;\;\;\;\;\;\;\;\;\;\;\;\;\;\;\;\;\;\; \ell = 1, \dots, L\\
\mathrm{\textbf{z}}_{\ell} &= \mathrm{MLP}(\mathrm{LN}(\mathrm{\textbf{z}}^{\prime}_{\ell})) + \mathrm{\textbf{z}}^{\prime}_{\ell},\;\;\;\;\;\;\;\;\;\;\;\;\;\;\;\;\;\;\;\;\;\;\;\;\;\;\;\;\;\;\;\;\;\;\;\;\; \ell = 1, \dots, L\\
\mathrm{\textbf{c}} &= \mathrm{LN}(\mathrm{\textbf{z}}^{0}_{L}),\\
\hat{y} &= \mathrm{L}(\mathrm{\textbf{c}}), 
\end{align}

\noindent where $\mathrm{MSA}$, $\mathrm{LN}$, $\mathrm{MLP}$, $\mathrm{L}$, $\mathrm{\textbf{E}}$, and $\mathrm{\textbf{E}}_{pos}$ are multi-head self-attention, layernorm,  multi-layer perceptron block (\textit{MLP}), linear layer, tile embedding projection, and position embedding (for more information see \cite{dosovitskiy2020image}). The variables $\mathrm{\textbf{E}}$ and $\mathrm{\textbf{E}}_{pos}$ are learnable. The layernorm applies normalization over a minibatch of inputs. In layernorm, the statistics are calculated independently across feature dimensions for each instance (i.e., tile) in a sequence (i.e., a bag of tiles). The multi-layer perceptron block is made of two linear layers followed by a dropout layer. The first linear layer has GELU activation function \cite{hendrycks2016gaussian}. The embedding is projected to a higher dimension in the first layer and then mapped to its original size in the second layer. Fig. \ref{fig:mlp} shows the structure of a \textit{MLP} block in a Transformer Encoder.

The remaining internal embeddings are passed to a dropout layer followed by a 1D convolution layer for the gene prediction head. The gene prediction head uses a dropout layer and 1D convolution layer as the output layer similar to the HE2RNA model introduced in \cite{schmauch2020deep}. However, the first two layers, which were two 1D convolution layers responsible for feature extraction in HE2RNA, were replaced with a Transformer Encoder to capture the relationship between all instances. As the model produces one prediction per gene per instance, the same aggregation strategy described in \cite{schmauch2020deep} was adapted for computing the gene prediction for each WSI. In particular, Schmauch et al. sampled a random number $n$ at each iteration and calculated each gene's prediction by averaging the top-$n$ predictions by tiles in a WSI (bag) \cite{schmauch2020deep}. They suggested this approach acts as a regularization technique and decreases the chance of overfitting \cite{schmauch2020deep}. As there were 49 tile embeddings in each bag,  $n$ was randomly selected from $\{1,\;2,\;5,\;10,\;20,\;49\}$. For a randomly selected $n$ during training, gene prediction outcome can be written as

\begin{align}
    \mathrm{\textbf{s}} &= \mathrm{Conv1D}(\mathrm{\textbf{z}}^{1:\textrm{end}}_{L}),\\
    \mathrm{\textbf{S}(n)} &= \sum_{i=1}^{n} \frac{\mathrm{\textbf{s}}^i}{n},
\end{align}

\noindent where $\mathrm{\textbf{z}}^{1:\textrm{end}}_{L} \in \mathbb{R}^{D \times k}$, $\mathrm{\textbf{s}} \in \mathbb{R}^{D \times k}$, and $\mathrm{\textbf{S}(n)} \in \mathbb{R}^{d_{g}}$ are the internal embeddings excluding the class token, the tile-wise gene prediction, and slide-level gene expression prediction, respectively. During the test the final prediction $\mathrm{\textbf{S}}$ is calculated as an average of all possible values for $n$ as

\begin{equation}
    \mathrm{\textbf{S}} = \sum_{i=1}^{k} \frac{\mathrm{\textbf{S}(i)}}{i}.
\end{equation}

\noindent The mean squared error loss function is employed to learn gene predictions.

\begin{figure}[ht]
 \centering
 \begin{subfigure}[b]{1\textwidth}
     \centering
     \includegraphics[width=1\textwidth]{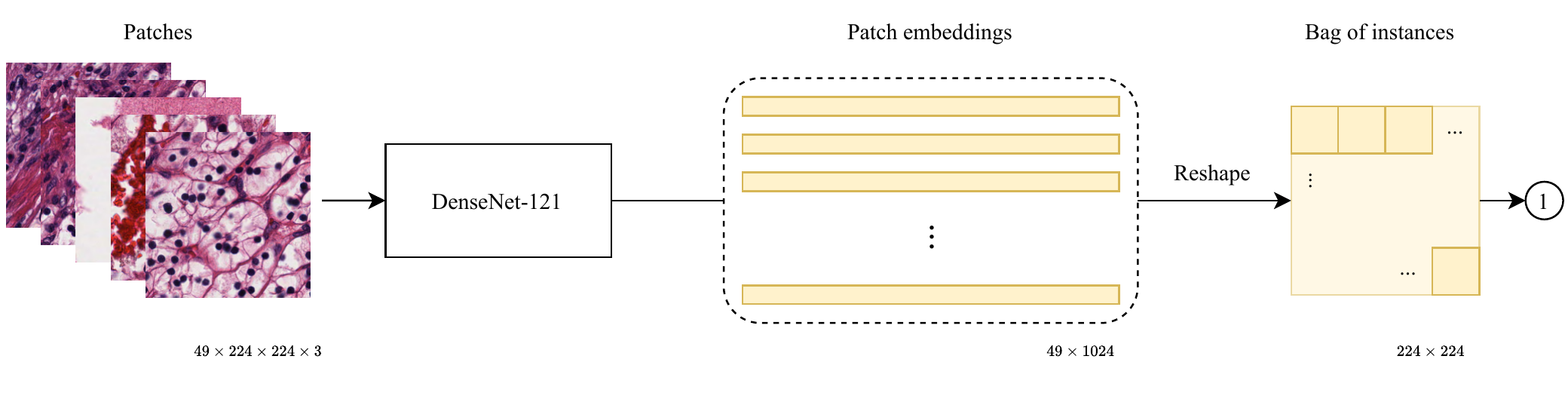}
     \caption{Creating a bag of instances from a WSI.}
 \end{subfigure}\\
 \begin{subfigure}[b]{1\textwidth}
     \centering
     \includegraphics[width=1\textwidth]{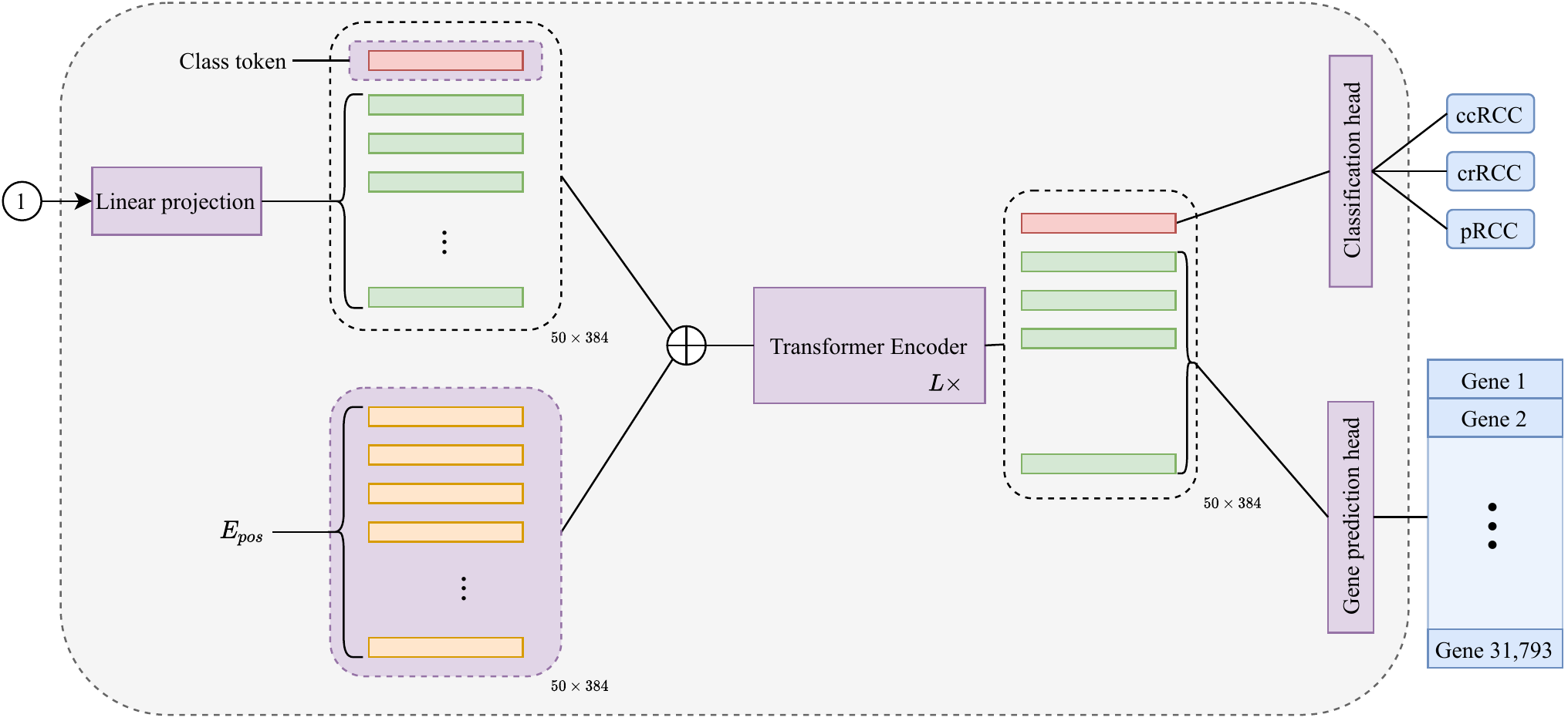}
     \caption{Internal schematic of how data flows in \textbf{tRNAsformer}.}
 \end{subfigure}
    \caption{A diagram showing how \textbf{tRNAsformer} works. (a) 49 tiles of size $224 \times 224 \times 3$ selected from 49 spatial clusters in a WSI are embedded with a DenseNet-121. The outcome is a matrix of size $49 \times 1024$ as DenseNet-121 has 1024 deep features after the last pooling. Then the matrix is reshaped and rearranged to $224 \times 224$ matrix in which each $32 \times 32$ block corresponds to a tile embedding $1 \times 1024$. (b) applying a 2D convolution with kernel 32, stride 32, and 384 kernels, each $32 \times 32$ block has linearly mapped a vector of 384 dimensional. Next, a class token is concatenated with the rest of the tile embeddings, and $\mathrm{\textbf{E}}_{pos}$ is added to the matrix before entering $L$ Encoder layers. The first row of the outcome, which is associated with the class token, is fed to the classification head. The rest of the internal embeddings that are associated with all tile embeddings are passed to the gene prediction head. All parts with learnable variables are shown in purple.}\label{fig:method}
    \label{fig:trnasformer}
\end{figure}

Finally, the total loss for \textbf{tRNAsformer} is computed as

\begin{align}
    \mathcal{L}_{\textrm{Total}}(\theta) &=  \mathcal{L}_{\textrm{classification}}(\theta) + \gamma \mathcal{L}_{\textrm{prediction}}(\theta) + \lambda \mathcal{L}_{\textrm{regularization}}(\theta),\\
    &= \frac{1}{B} \sum_{i=1}^{B} \left( - \mathrm{\textbf{y}}_i \mathrm{\textbf{log}} (\hat{y}_i) + \gamma |\mathrm{\textbf{y}}_i^g - S_i| \right) + \lambda {|\theta|}^2_2,
\end{align}

\noindent where $\theta$, $\lambda$, $\gamma$, $B$, $\mathrm{\textbf{y}}^g$ are the model parameters, weight regularization coefficient, hyperparameter for scaling the losses, number of samples in a batch, and true bulk RNA-seq associated with the slides. A summary of the proposed approach is included in Fig. \ref{fig:method}.

\paragraph{Training settings for training \textbf{tRNAsformer} models.} To begin, TCGA cases are split into $80\%$, $10\%$, and $10\%$ subsets for the training, validation, and test sets. Each case was associated with a patient and could have contained multiple diagnostic WSIs or RNA-seq files. The 100 bags were sampled from each WSI. As a result, the training set comprised of 63,400 bags (see Table \ref{tab:tcga-kidney}). 

The \textbf{tRNAsformer}'s internal representation size was set to 384. The \textit{MLP} ratio and the number of self-attention heads were both four. The \textbf{tRNAsformer} was trained for 20 epochs with a minibatch of size 64. The AdamW was chosen as the optimizer with a starting learning rate of $3 \times 10^{-4}$ \cite{loshchilov2017decoupled}. The weight regularization coefficient was set to 0.01 to avoid overfitting. The reduce-on-plateau method was chosen for scheduling the learning rate. Therefore, the learning rate was reduced by ten every two epochs without an improvement in the validation loss. The scaling coefficient $\gamma$ was set to 0.5. The last dropout layer's probability was set to 0.25. The values for the model with the lowest validation loss are reported. All experiments are conducted using a single NVIDIA GeForce RTX 2080 SUPER graphic card. The desktop's CPU was Intel(R) Core(TM) i9-10900X.

\paragraph{Training settings for training MLP model.} Another model was trained based on the MLP architecture described in \cite{schmauch2020deep} for fair comparison. The fully connected layers were replaced with successive 1D convolutions with kernel size one and stride one to slide data due to practicality in the MLP design \cite{schmauch2020deep}. A dropout layer is applied between successive layers, and the activation function was ReLU. The model based on MLP design suggested in \cite{schmauch2020deep} is referred to as $\textrm{HE2RNA}_{\textrm{bb}}$\footnote{bb stands for \textbf{b}ack\textbf{b}one} as it was trained on TCGA training set used in this dissertation. The $\textrm{HE2RNA}_{\textrm{bb}}$ model is made of three 1D convolutional layers. The first two layers each contained $h$ input and output channels, whereas the last layer had the same number of output channels as the number of genes. In other words, $h$ is the size of the model's internal representation. The $h$ was set to 1024 for $\textrm{HE2RNA}_{\textrm{bb}1024}$. The model was trained for 20 epochs using AdamW optimizer and a starting learning rate of $3 \times 10^{-4}$ \cite{loshchilov2017decoupled}. If no improvement is observed for the validation loss for two epochs, the learning rate was reduced by ten. The minibatch size was set to 64. The values for the model with the lowest validation loss are provided. The number of parameters of each model in Table \ref{tab:rna-model-parameters} for comparison. The wall clock time for a single epoch for training and validation is also provided in the same table as the number of parameters.

\paragraph{The Ohio State University kidney dataset.}\label{sec:external-dataset} This is an internal dataset that we used to evaluate the internal representation of our model. The pathology department's surgical pathology files were examined for consecutive cases of renal cell carcinoma classified as clear cell carcinoma (ccRCC), chromophobe renal cell carcinoma (crRCC), or papillary renal cell carcinoma (pRCC). The dataset was created at the end of the search, and it contained 142 instances of renal cell carcinoma. The WSIs from ccRCC, crRCC, and pRCC were 48, 44, and 50, respectively. Each patient had one representative cancer slide that was examined by a board certified pathologist (Anil V. Parwani) before being scanned at $20 \times$ utilising an aperio XT scanscope (Leica biosystems, CA). A board-certified pathologist (AP) reviewed the WSI images and validated the classifications a second time to guarantee the image quality and correctness of the diagnosis.

\paragraph{External validation of the transcriptomic learning for representing WSIs.} The model that was trained on the TCGA kidney dataset was used to embed the external dataset. The classification and WSI search studies were then performed to examine domain change impact on the proposed pipeline.

\section*{Author contributions}

A.S. designed and performed the research, analyzed and interpreted the results, and wrote the paper. HR.T. and J.D.H. have conceived and oversaw the study. All authors read and approved the final manuscript.

\section*{Competing interests}
All authors have no conflict of interests and nothing to declare. 

\section*{Data availability}
The NCI Genomic Data Commons Portal (https://portal.gdc.cancer.gov/) has all of the TCGA digital slides available to the public. 

\section*{Code availability}
Upon publishing, our source codes will be made publicly available on our lab's website: ``kimia.uwaterloo.ca''.

\section*{Declarations}

\subsection*{Ethics approval and consent to participate}

This study was approved by the Ohio State University institutional research board. Informed written consent was obtained from all individual patients included in the study. All methods were carried out in accordance with relevant guidelines and regulations. All the data was de-identified using an honest broker system.

\subsection*{Consent for Publication}
Not applicable.

\subsection*{Funding}
This project was partially funded as part of an ORF-RE consortium by the Government of Ontario.

\subsection*{Acknowledgements}
Not applicable.

\bibliographystyle{unsrt}  
\bibliography{references}  

\appendix

\newpage

\section{Appendix}

\begin{table}[ht]
  \caption{TCGA kidney dataset split for transcriptomic learning (the number of cases, slides, and FPKM files per subtype per subset).}
  \label{tab:tcga-kidney}
  \centering
  \begin{tabular}{lccccccccc}
    \toprule
    &   \multicolumn{3}{c}{Train}   &   \multicolumn{3}{c}{Validation}  &   \multicolumn{3}{c}{Test}  \\
    \cmidrule(lr){2-4}   \cmidrule(lr){5-7} \cmidrule(lr){8-10}
    Subtype &   Cases   &   Slides  &   FPKMs &   Cases   &   Slides  &   FPKMs &   Cases   &   Slides  &   FPKMs    \\
    \midrule
    ccRCC &   369 &   372 &   373 &   43  &   43  &   45  &   46  &   47  &   48  \\
    crRCC &   47  &   47  &   47  &   8   &   8   &   8   &   7   &   7   &   7   \\
    pRCC  &   195 &   215 &   195 &   26  &   28  &   26  &   24  &   26  &   24  \\
    \bottomrule
  \end{tabular}
\end{table}

\begin{table}[ht]
    \centering
    \caption{The number of parameters and one epoch's wall clock processing time for \textbf{tRNAsformer} and $\textrm{HE2RNA}_{\textrm{bb}}$ models. When the minibatch is set to 64, the processing time is the wall clock time for one epoch of training or validation.}\label{tab:rna-model-parameters}
    \begin{tabular}{lccccccc}
    \hline
    & & \multicolumn{2}{c}{Processing time (s)} \\
    \cmidrule(lr){3-4}
    Model & Number of parameters    &   Training    &   Validation\\
    \hline
    $\textrm{tRNAsformer}_{L=1}$  & 14,429,876  &   128   &   61  \\
    $\textrm{tRNAsformer}_{L=2}$  & 16,204,340  &   133   &   60  \\
    $\textrm{tRNAsformer}_{L=4}$  & 19,753,268  &   146   &   61  \\
    $\textrm{tRNAsformer}_{L=8}$  & 26,851,124  &   173   &   64  \\
    $\textrm{tRNAsformer}_{L=12}$  & 33,948,980 &   205   &   65 \\ \hline
    $\textrm{HE2RNA}_{\textrm{bb}1024}$   &   34,687,025    &   335   &   81    \\
    \hline
    \end{tabular}
\end{table}

\begin{figure}[ht]
  \centering
  \begin{subfigure}[b]{0.4\textwidth}
      \centering
      \includegraphics[width=0.6\linewidth]{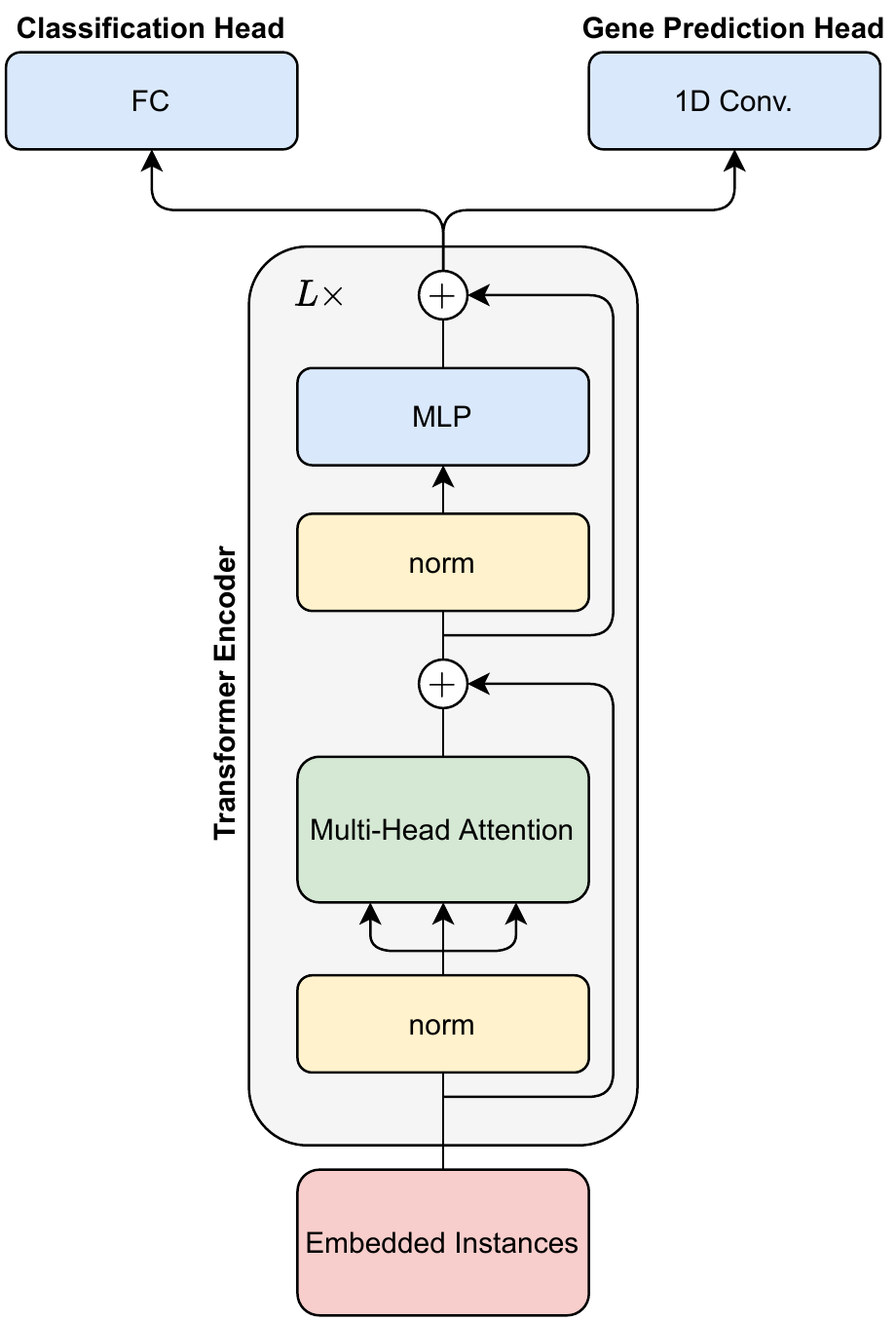}
      \caption{\textbf{tRNAsformer} model}
  \end{subfigure}
  \begin{subfigure}[b]{0.3\textwidth}
      \centering
      \includegraphics[width=0.5\linewidth]{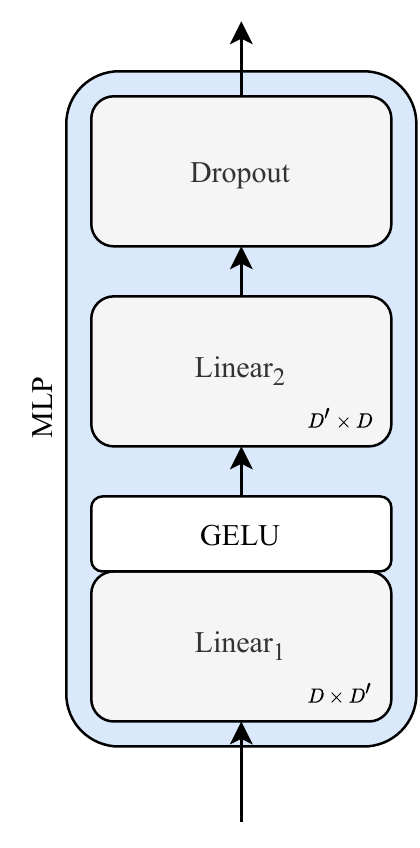}
      \caption{structure of the \textit{MLP} block}\label{fig:mlp}
  \end{subfigure}
  \caption{The \textbf{tRNAsformer} model architecture -- (a) a standard Transformer Encoder comprises layernorm, multi-head attention, multi-layer perceptron block, and residual skip connections. Because it is a multi-head self-attention module, the first layernorm's output embedding is provided to the multi-head attention as the query, key, and value. Each model can have $L$ blocks of Transformer Encoder. The classification head transforms the internal representation to the number of classes, whereas the gene prediction head maps it to the number of genes. (b) a detailed diagram of multi-layer perceptron block (\textit{MLP}). The letter $D$ refers to the size of internal representation in the Transformer Encoder, and $\frac{D^{\prime}}{D}$ is referred to as \textit{MLP} ratio.}
  \label{fig:diagram}
\end{figure}

\begin{figure}[ht]
    \centering
    \begin{subfigure}{0.4\textwidth}
    \includegraphics[width=1\linewidth]{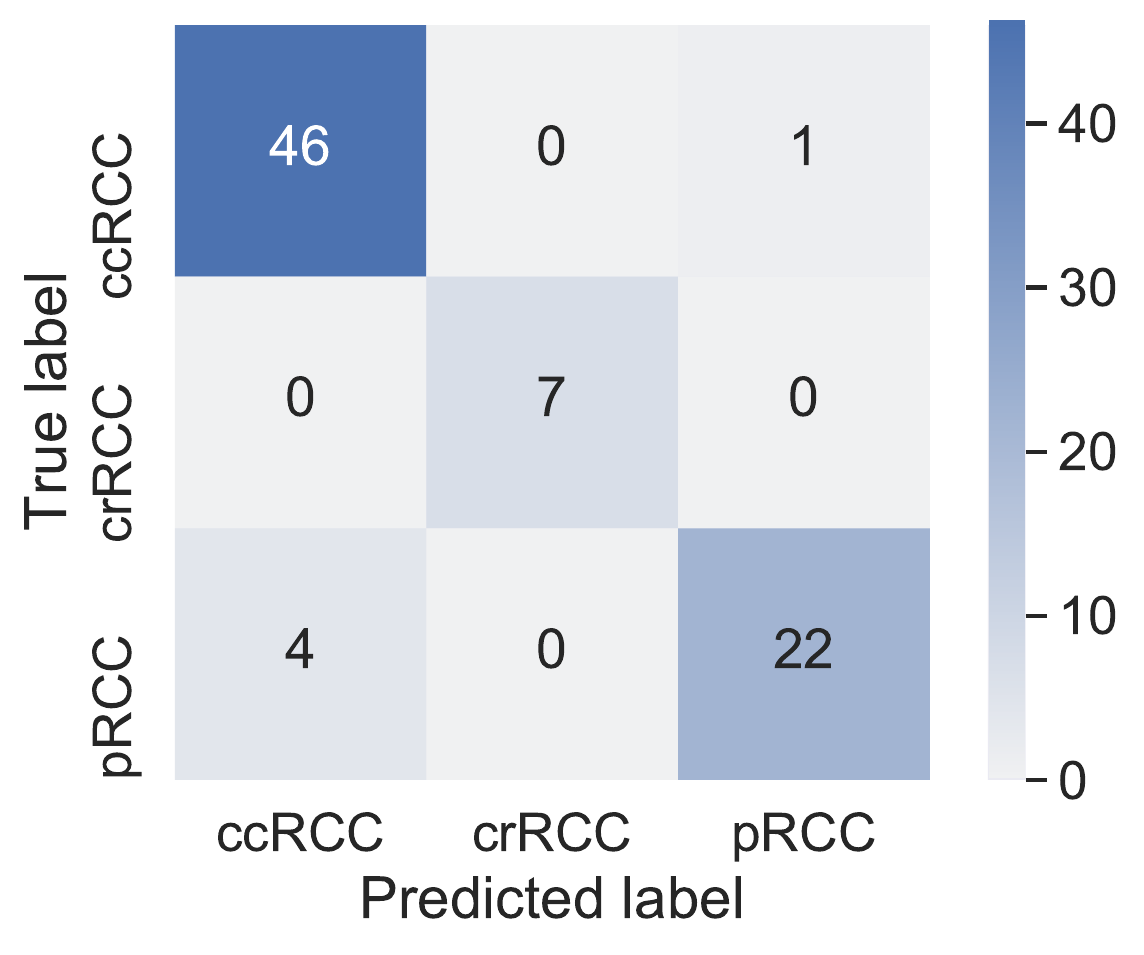}
    \caption{$\textrm{tRNAsformer}_{L=1}$}
    \end{subfigure}
    \begin{subfigure}{0.4\textwidth}
    \includegraphics[width=1\linewidth]{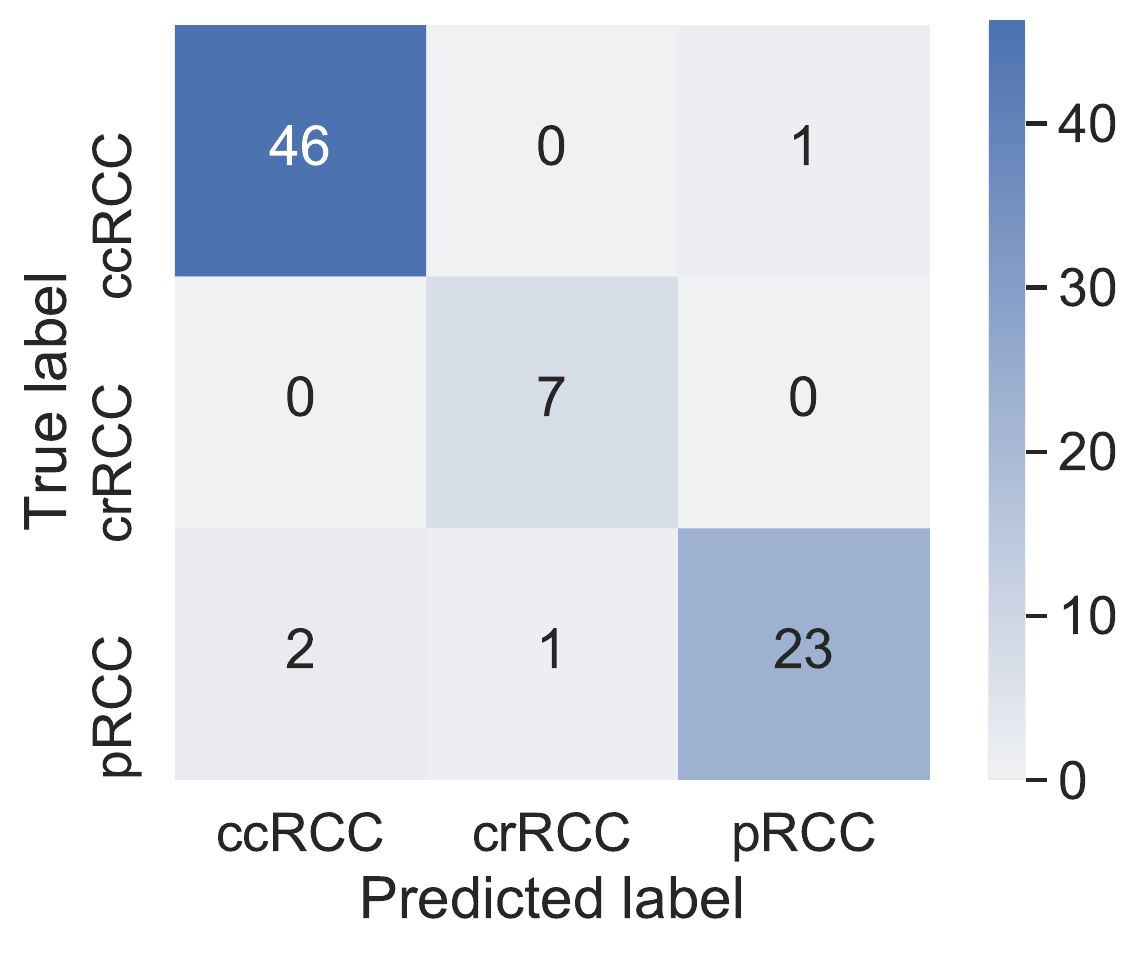}
    \caption{$\textrm{tRNAsformer}_{L=2}$}
    \end{subfigure}\\
    \begin{subfigure}{0.4\textwidth}
    \includegraphics[width=1\linewidth]{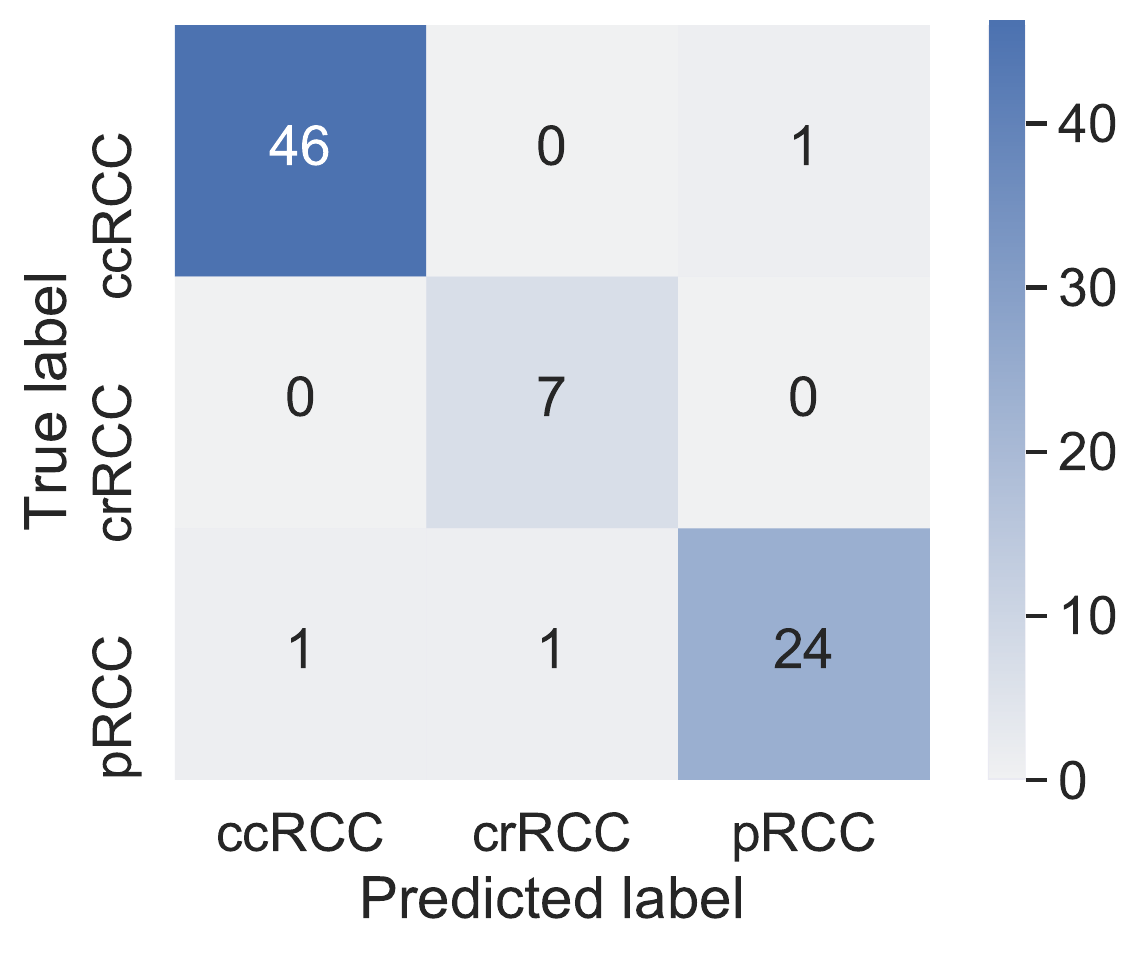}
    \caption{$\textrm{tRNAsformer}_{L=4}$}
    \end{subfigure}
    \begin{subfigure}{0.4\textwidth}
    \includegraphics[width=1\linewidth]{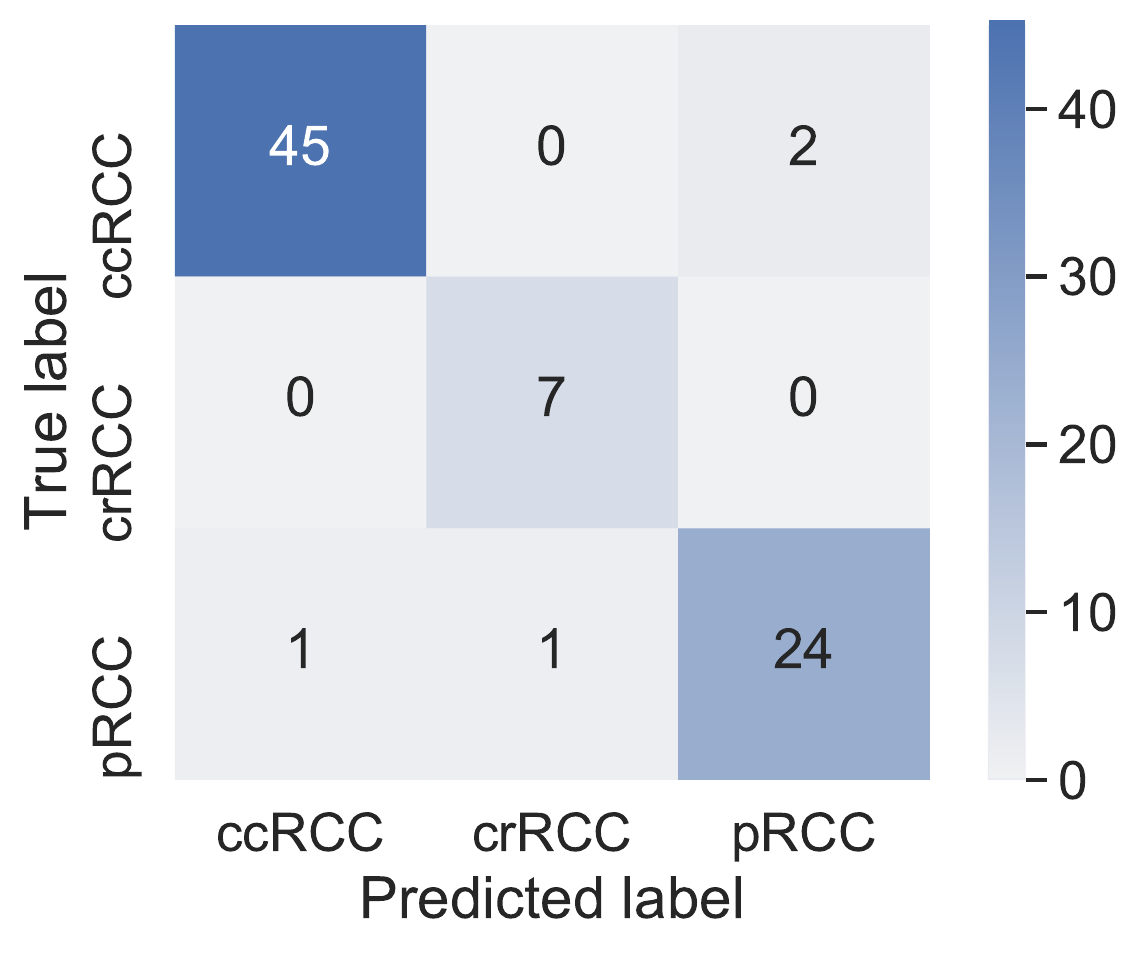}
    \caption{$\textrm{tRNAsformer}_{L=8}$}
    \end{subfigure}\\
    \begin{subfigure}{0.4\textwidth}
    \includegraphics[width=1\linewidth]{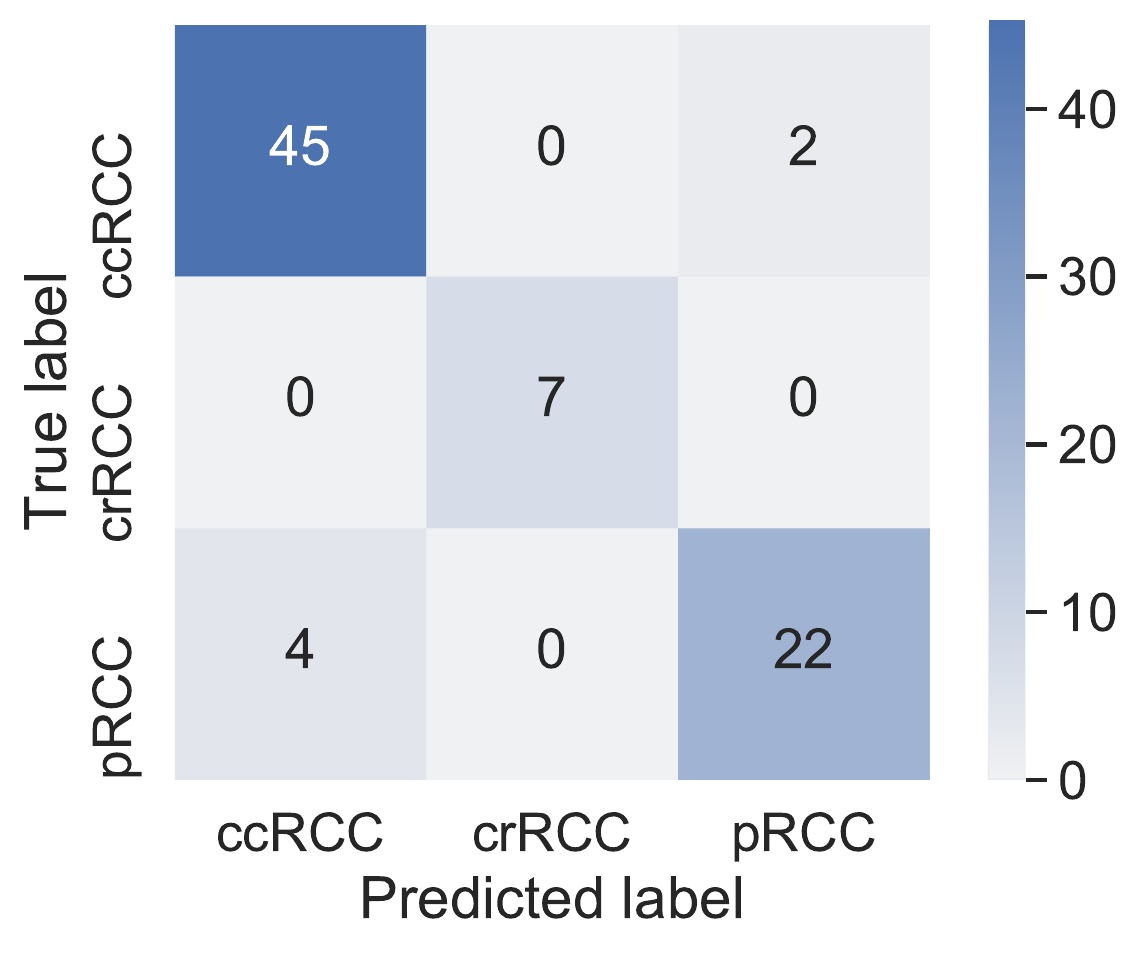}
    \caption{$\textrm{tRNAsformer}_{L=12}$}
    \end{subfigure}
    \caption{The confusion matrices for different models applied on 8,000 bags created from 80 TCGA test WSIs. (a)-(f) are for $\textrm{tRNAsformer}_{L}$, $L=(1,2,4,8,12)$, respectively.}
    \label{fig:rna-confusion-tcga}
\end{figure}

\begin{figure}
    \centering
    \begin{subfigure}{0.4\textwidth}
    \includegraphics[width=1\linewidth]{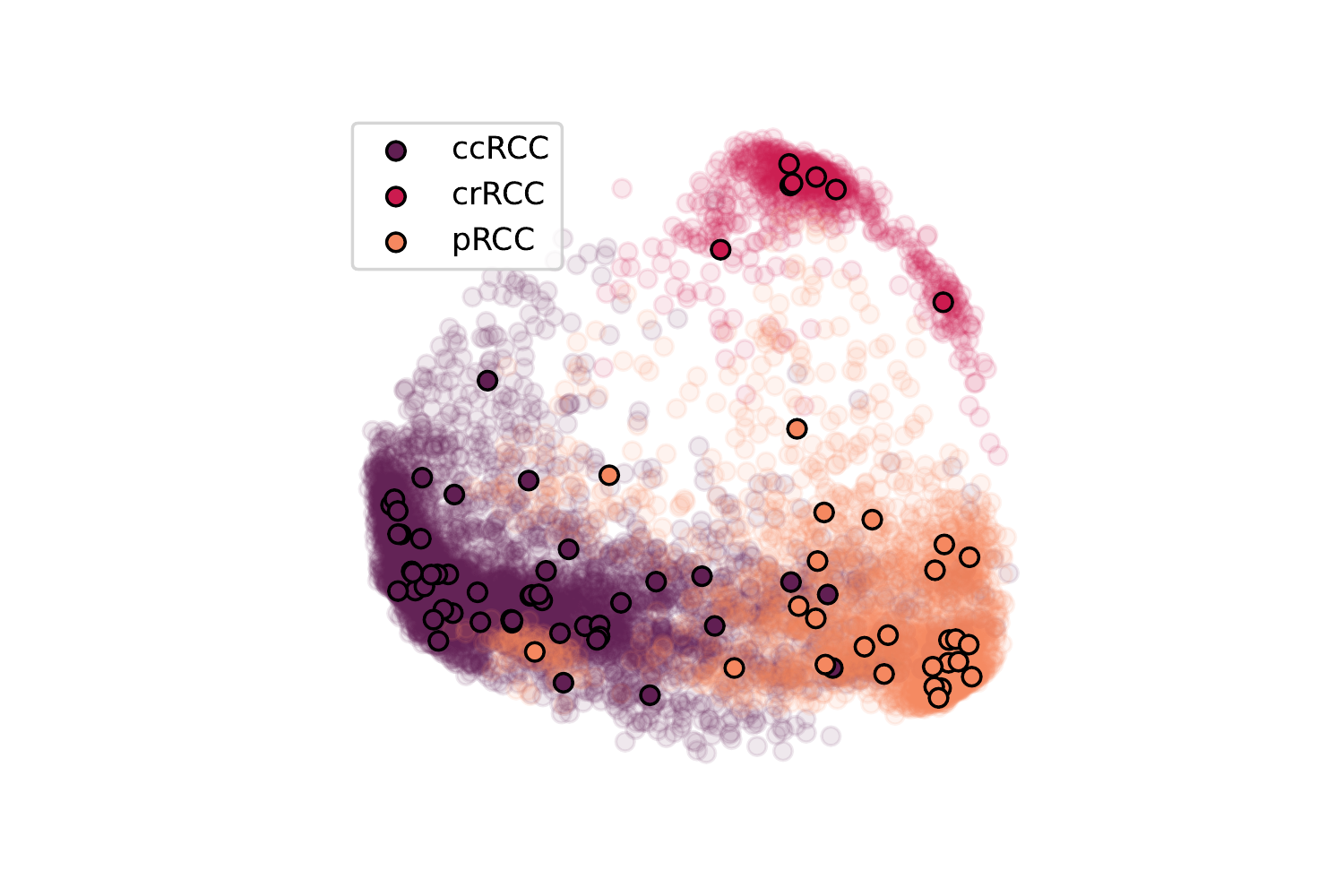}
    \caption{$\textrm{tRNAsformer}_{L=1}$}
    \end{subfigure}
    \begin{subfigure}{0.4\textwidth}
    \includegraphics[width=1\linewidth]{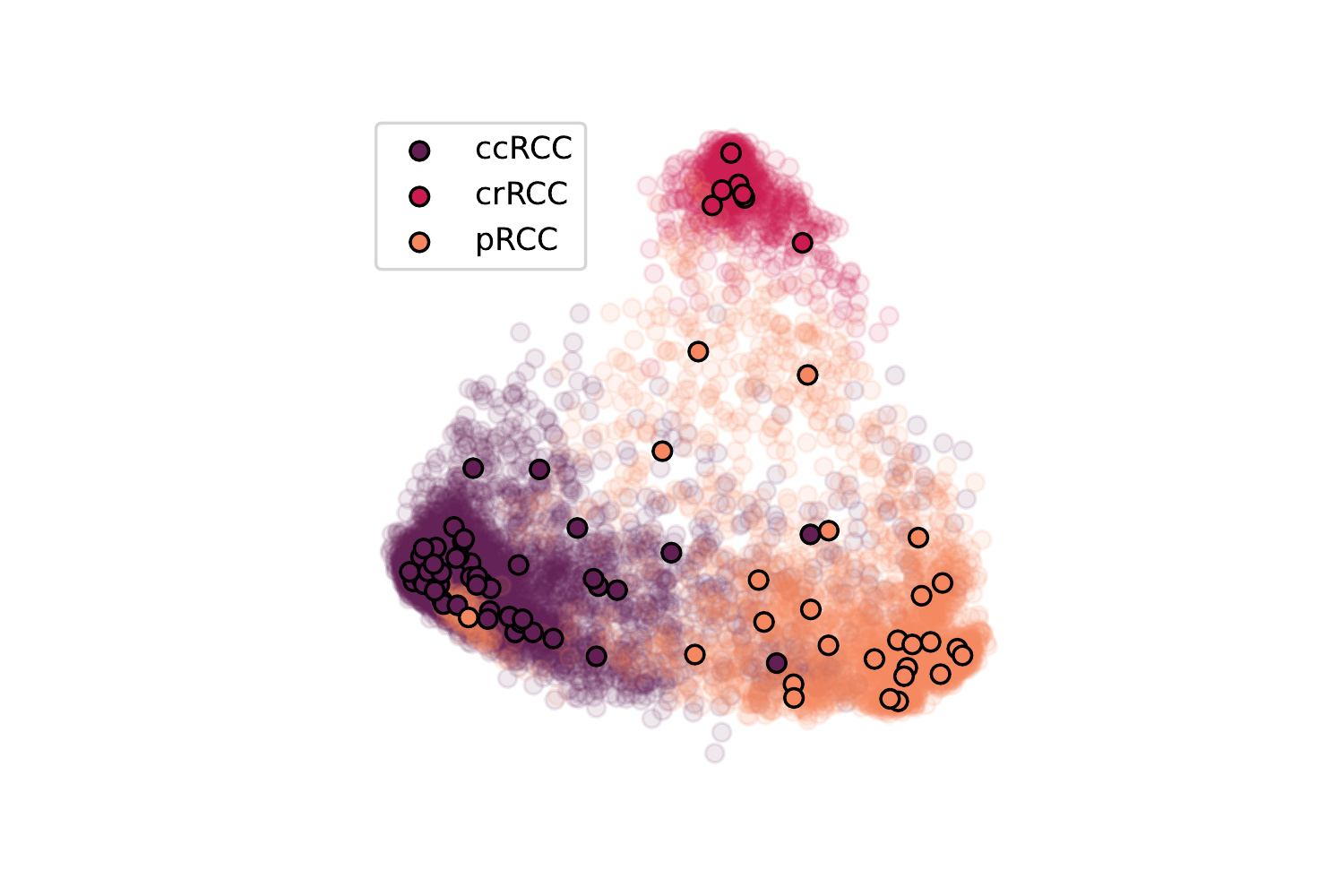}
    \caption{$\textrm{tRNAsformer}_{L=2}$}
    \end{subfigure}\\
    \begin{subfigure}{0.4\textwidth}
    \includegraphics[width=1\linewidth]{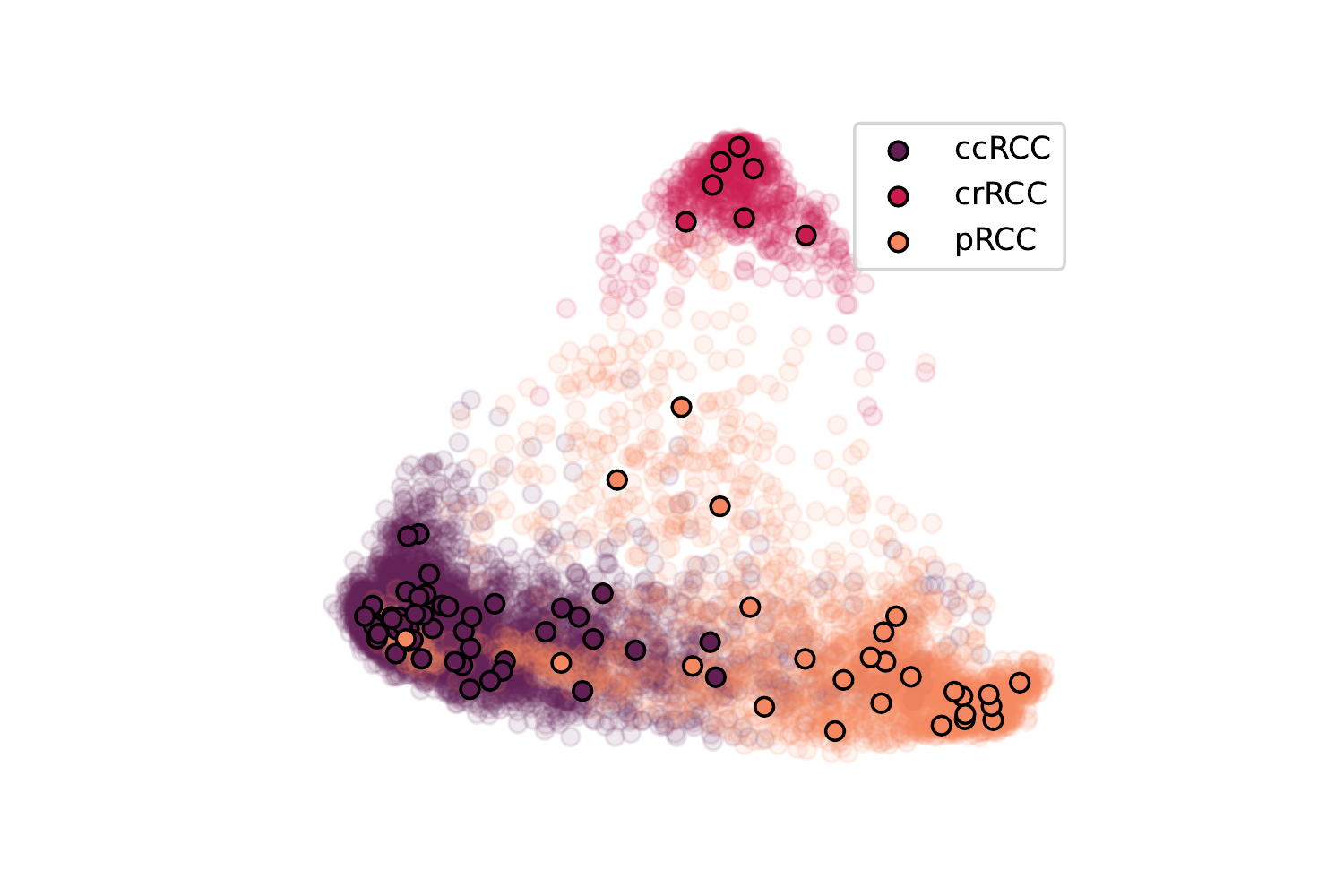}
    \caption{$\textrm{tRNAsformer}_{L=4}$}
    \end{subfigure}
    \begin{subfigure}{0.4\textwidth}
    \includegraphics[width=1\linewidth]{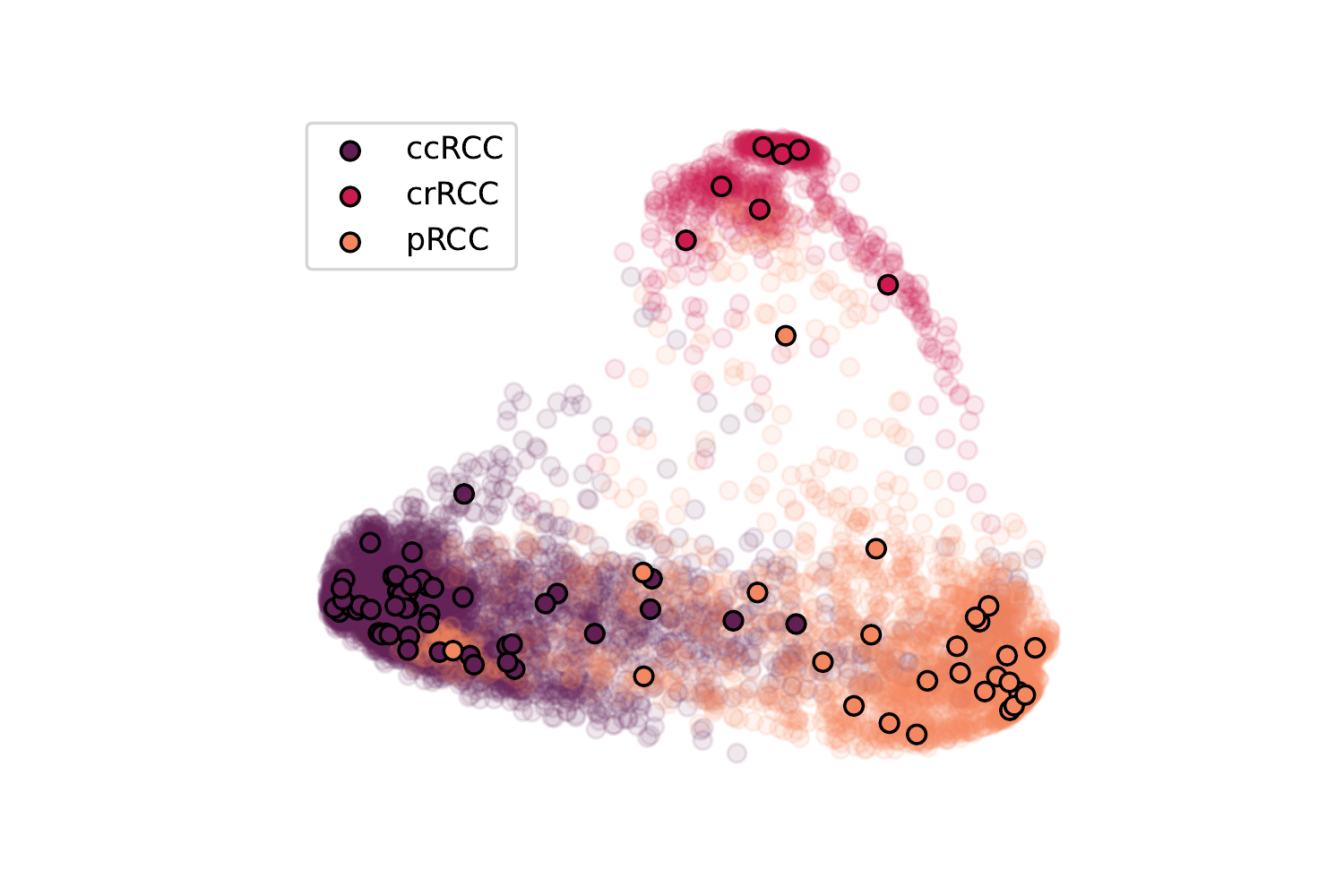}
    \caption{$\textrm{tRNAsformer}_{L=8}$}
    \end{subfigure}\\
    \begin{subfigure}{0.4\textwidth}
    \includegraphics[width=1\linewidth]{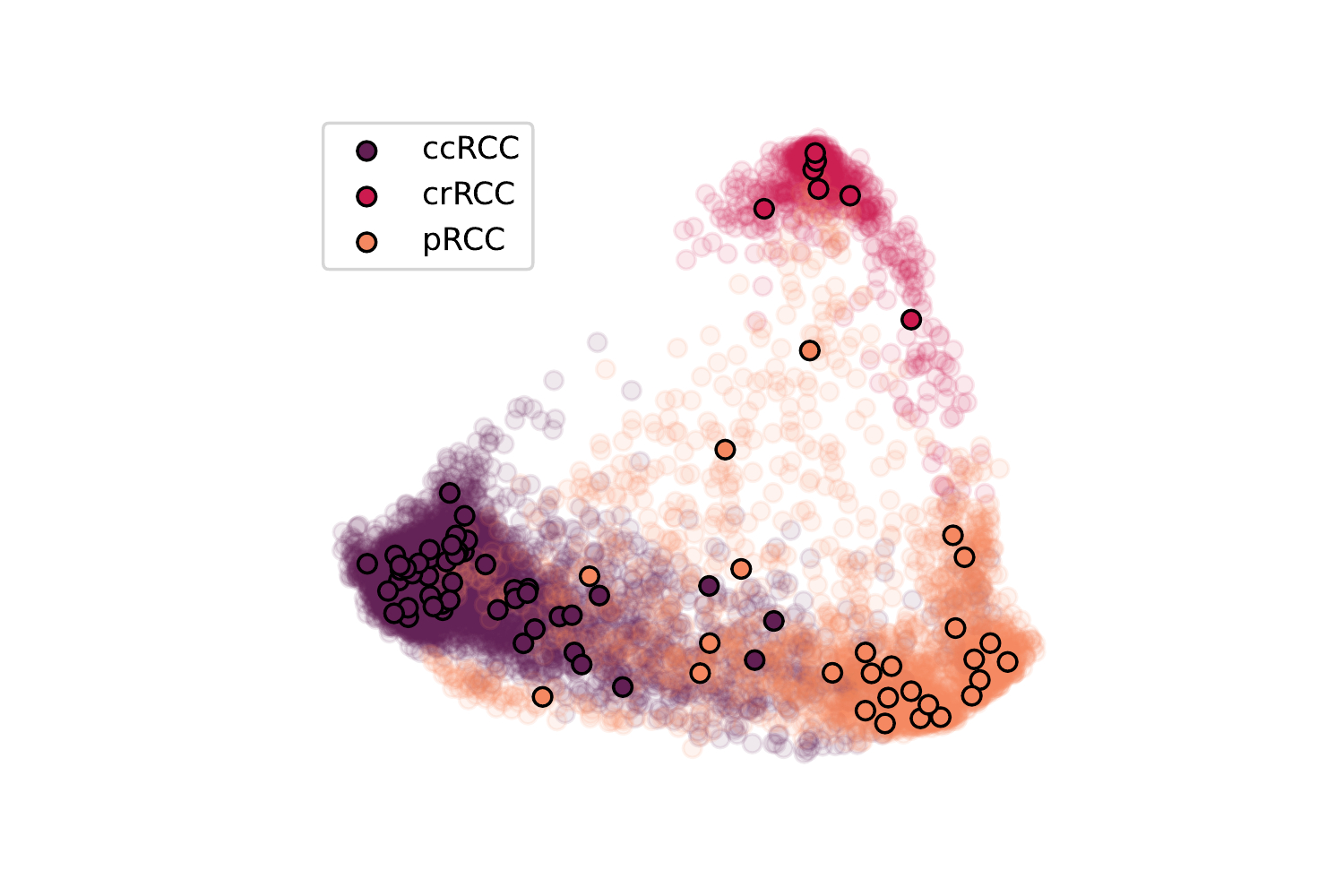}
    \caption{$\textrm{tRNAsformer}_{L=12}$}
    \end{subfigure}
    \caption{The two-dimensional PCA projection of TCGA test WSI features. (a)-(f) are for $\textrm{tRNAsformer}_{L}$, $L=(1,2,4,8,12)$, respectively. Each TCGA test WSI is represented by 100 bags of features. All bags of features associated with the test set are shown with transparent circles. The average of PCA projection of each WSI (average of 100 bags associated with each WSI) is shown in bold circles with black edges.}
    \label{fig:rna-pca-tcga}
\end{figure}

\begin{figure}
    \centering
    \begin{subfigure}{0.39\textwidth}
    \includegraphics[width=1\linewidth]{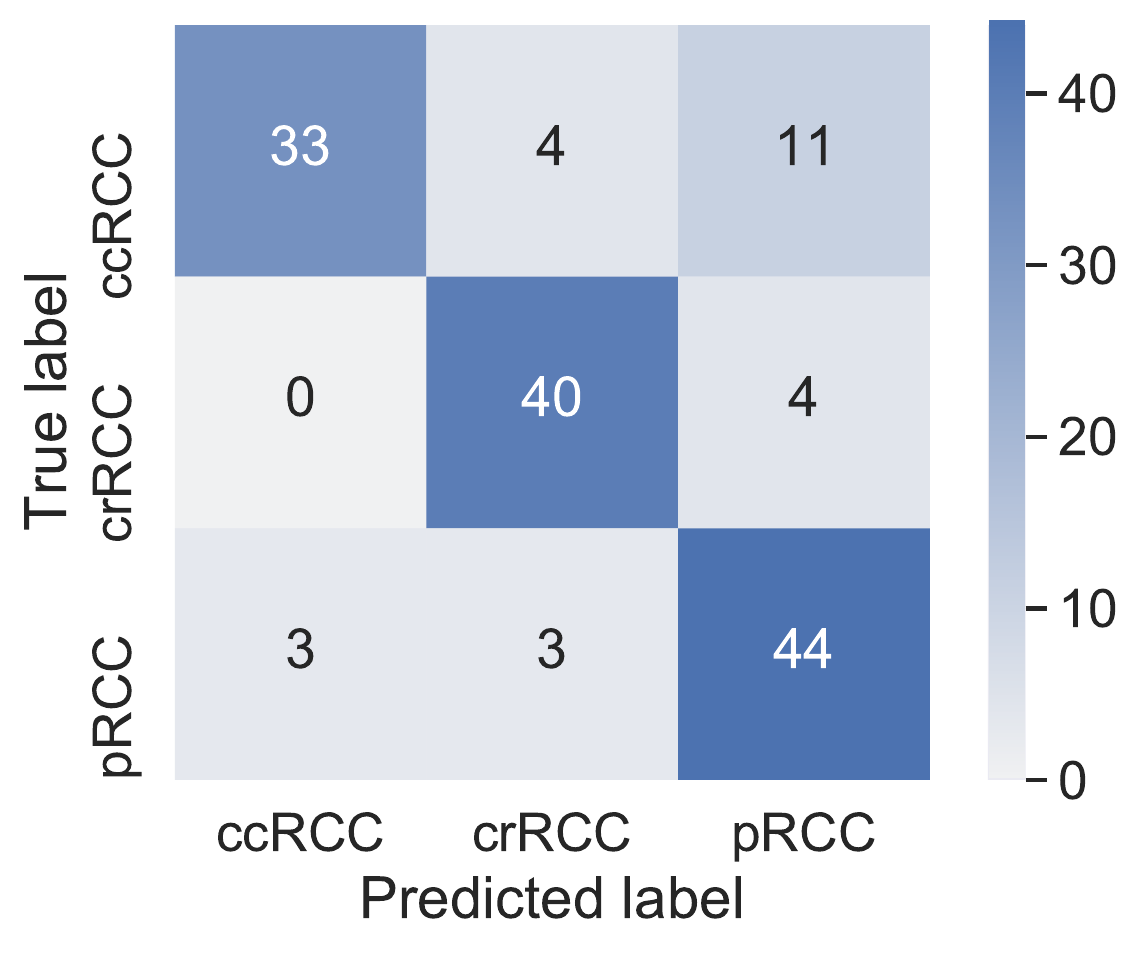}
    \caption{$\textrm{tRNAsformer}_{L=1}$}
    \end{subfigure}
    \begin{subfigure}{0.39\textwidth}
    \includegraphics[width=1\linewidth]{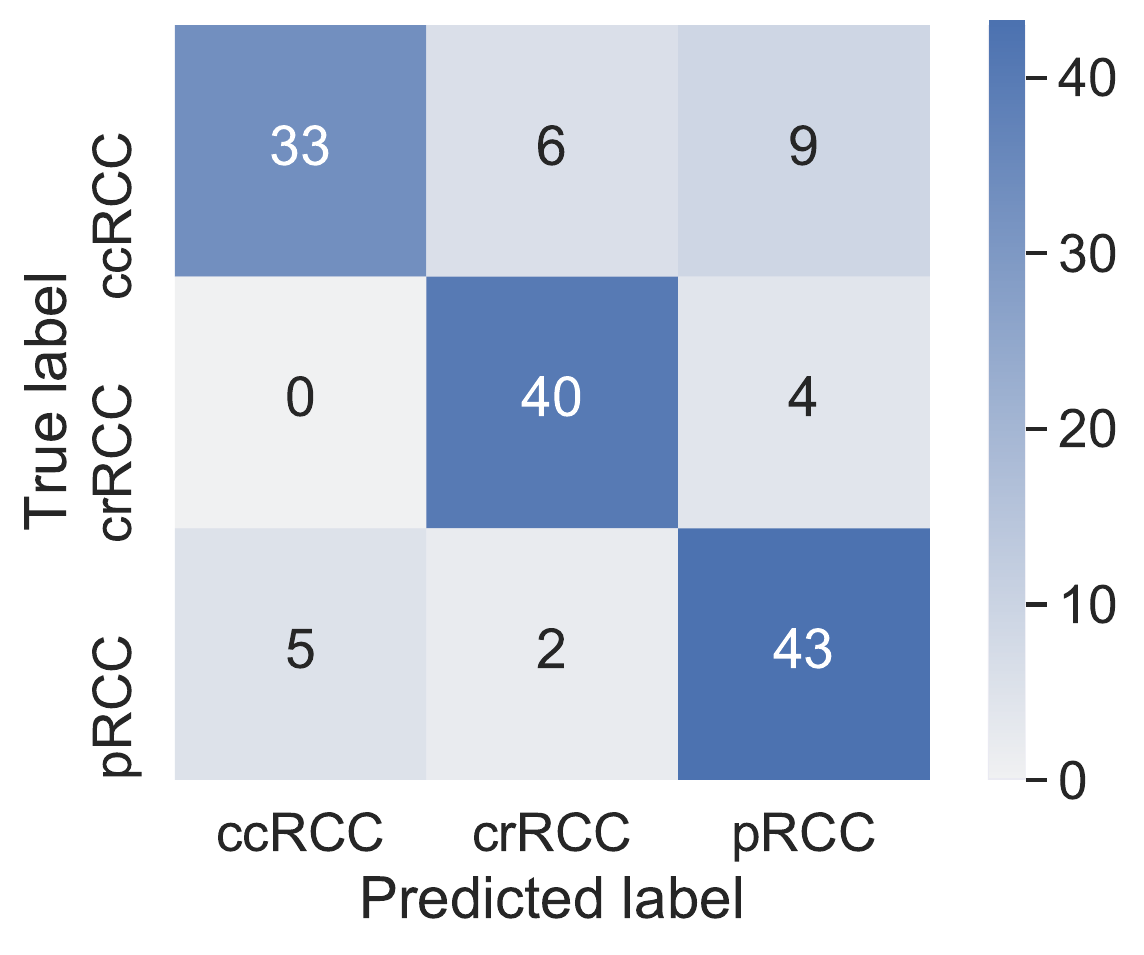}
    \caption{$\textrm{tRNAsformer}_{L=2}$}
    \end{subfigure}\\
    \begin{subfigure}{0.39\textwidth}
    \includegraphics[width=1\linewidth]{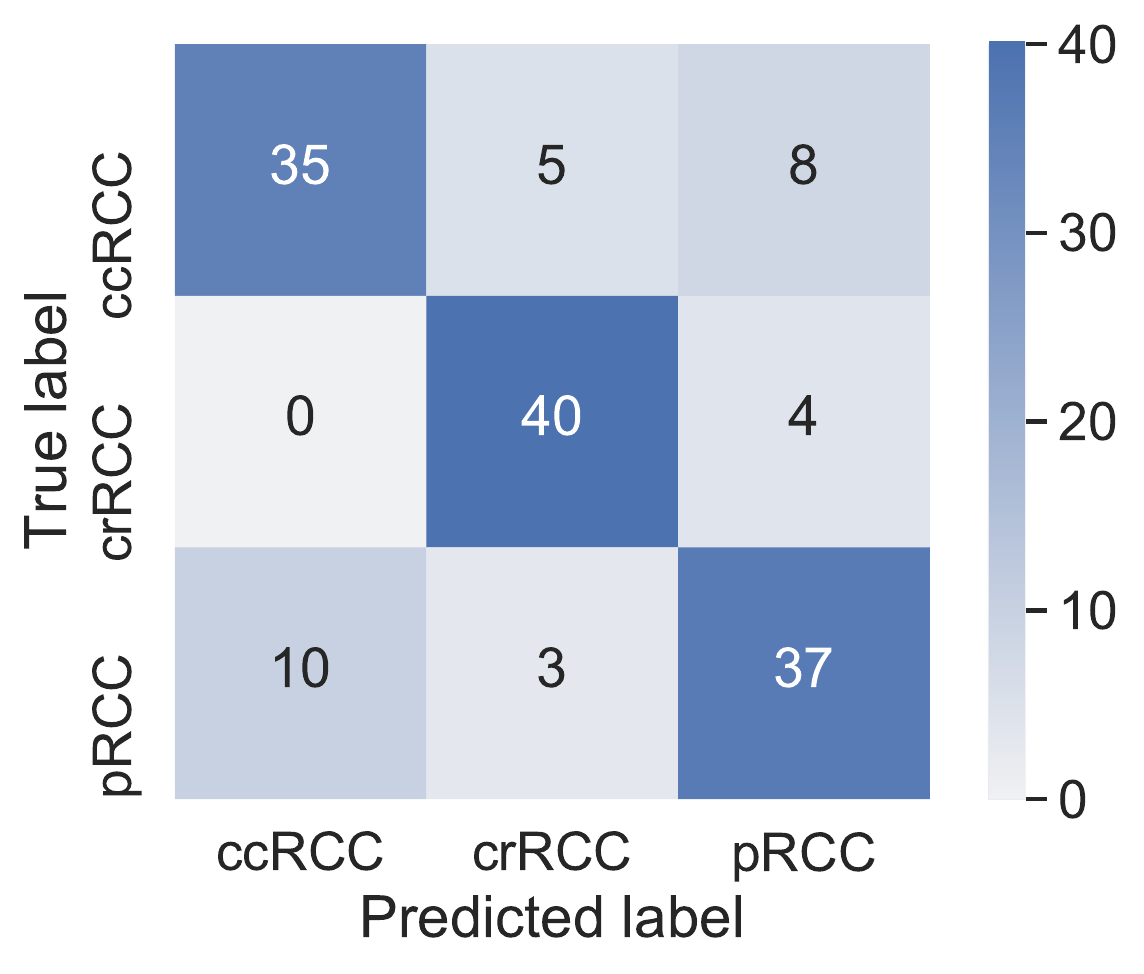}
    \caption{$\textrm{tRNAsformer}_{L=4}$}
    \end{subfigure}
    \begin{subfigure}{0.39\textwidth}
    \includegraphics[width=1\linewidth]{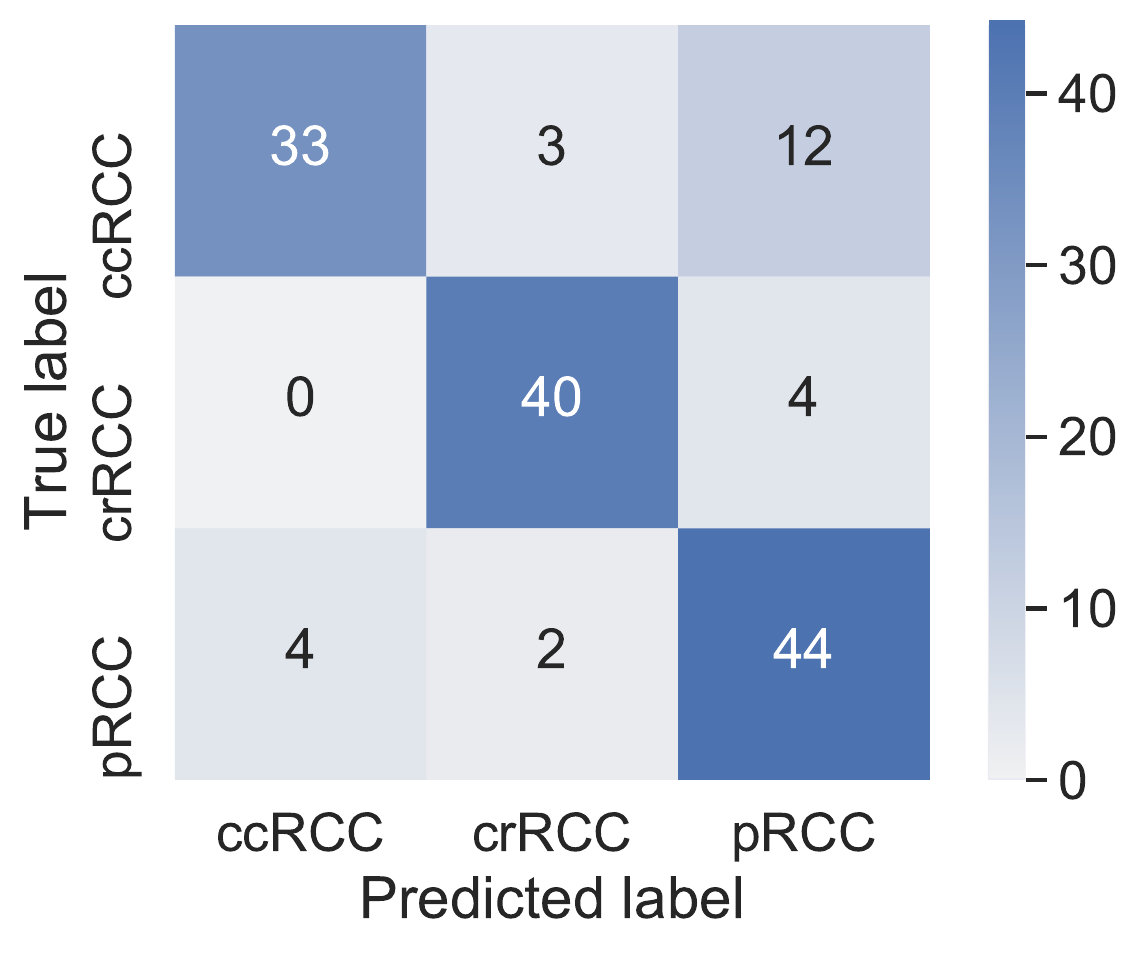}
    \caption{$\textrm{tRNAsformer}_{L=8}$}
    \end{subfigure}
    \begin{subfigure}{0.39\textwidth}
    \includegraphics[width=1\linewidth]{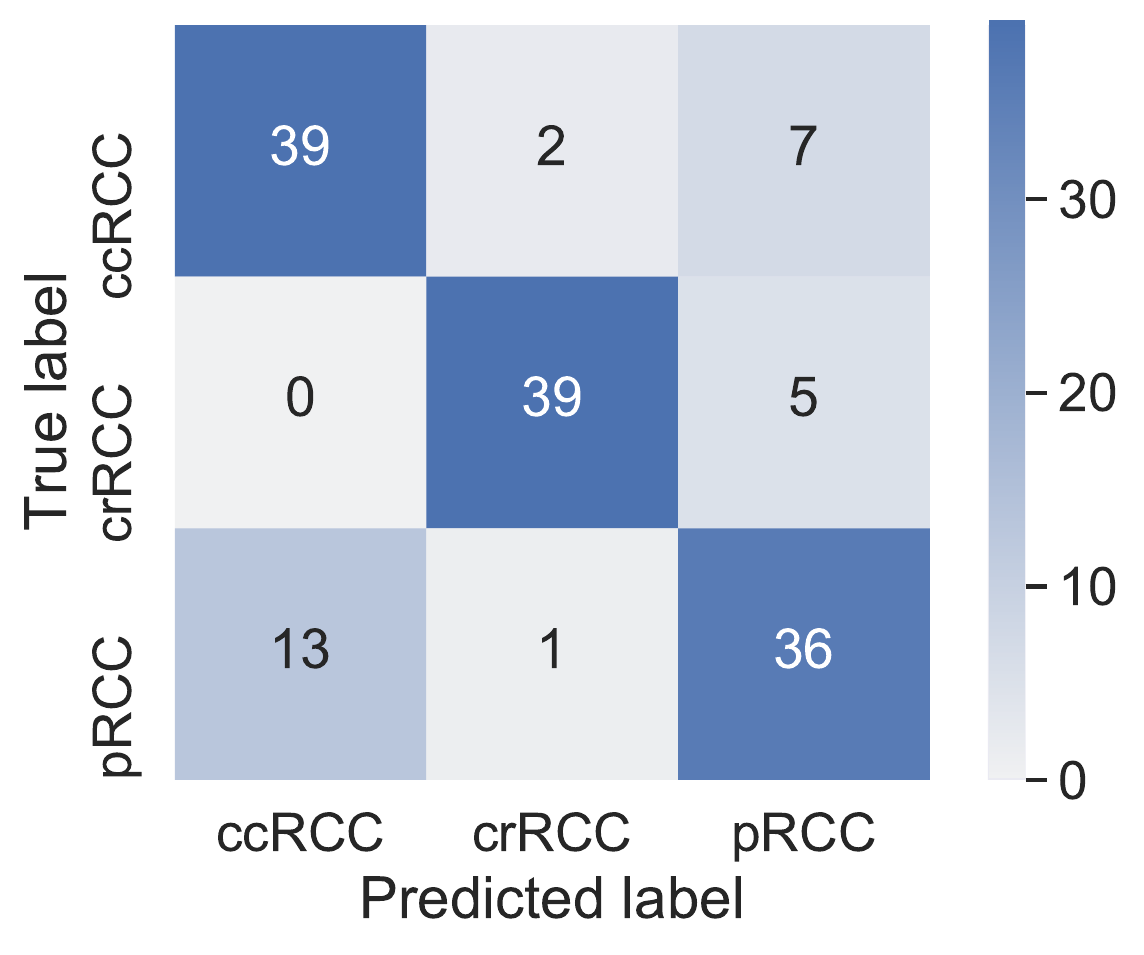}
    \caption{$\textrm{tRNAsformer}_{L=12}$}
    \end{subfigure}
    \caption{The confusion matrices for different models applied on 14,200 bags created from the external dataset WSIs. (a)-(d) are for $\textrm{tRNAsformer}_{L}$, $L=(1,2,4,8,12)$, respectively.}
    \label{fig:rna-confusion-external}
\end{figure}

\begin{figure}
    \centering
    \begin{subfigure}{0.4\textwidth}
    \includegraphics[width=1\linewidth]{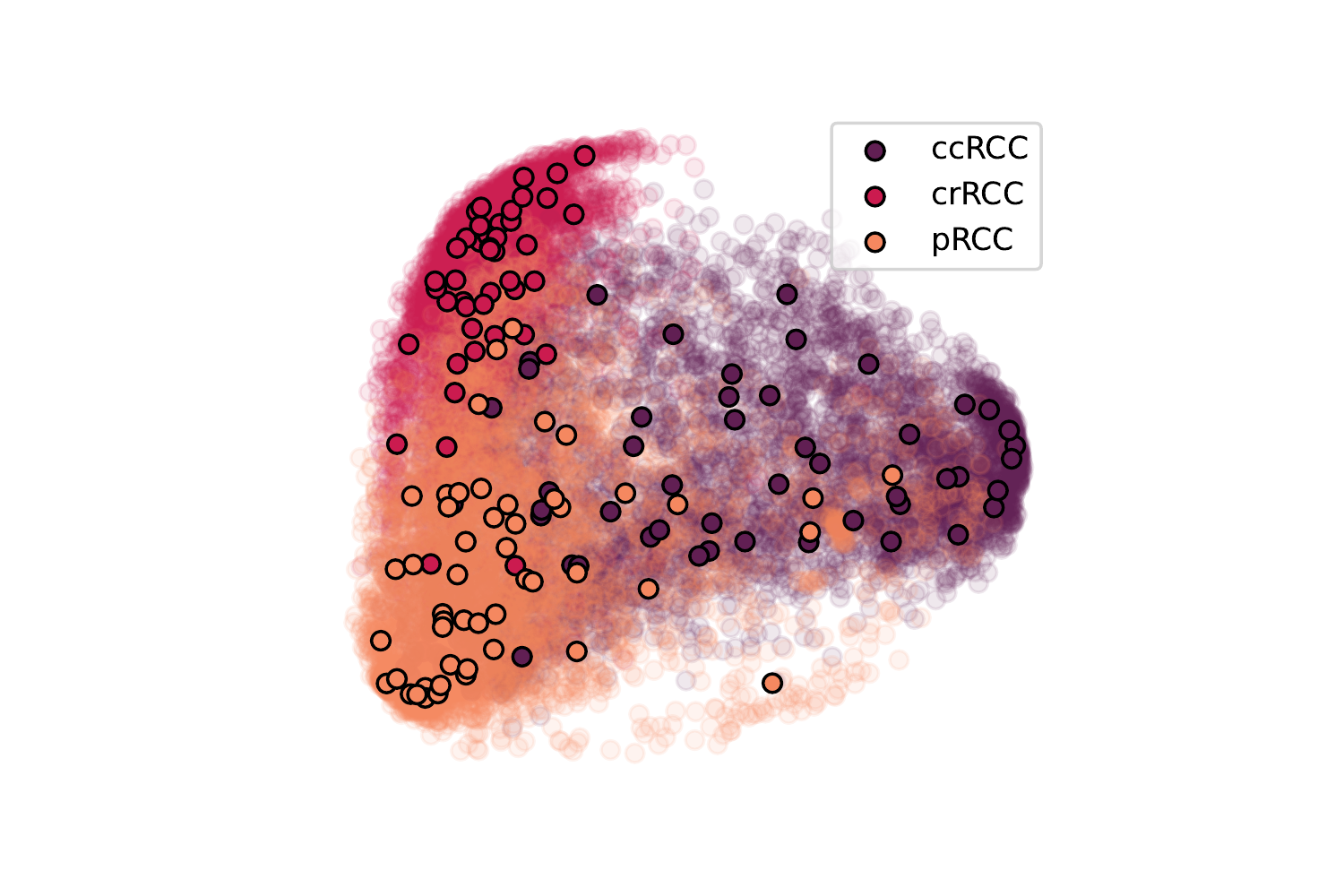}
    \caption{$\textrm{tRNAsformer}_{L=1}$}
    \end{subfigure}
    \begin{subfigure}{0.4\textwidth}
    \includegraphics[width=1\linewidth]{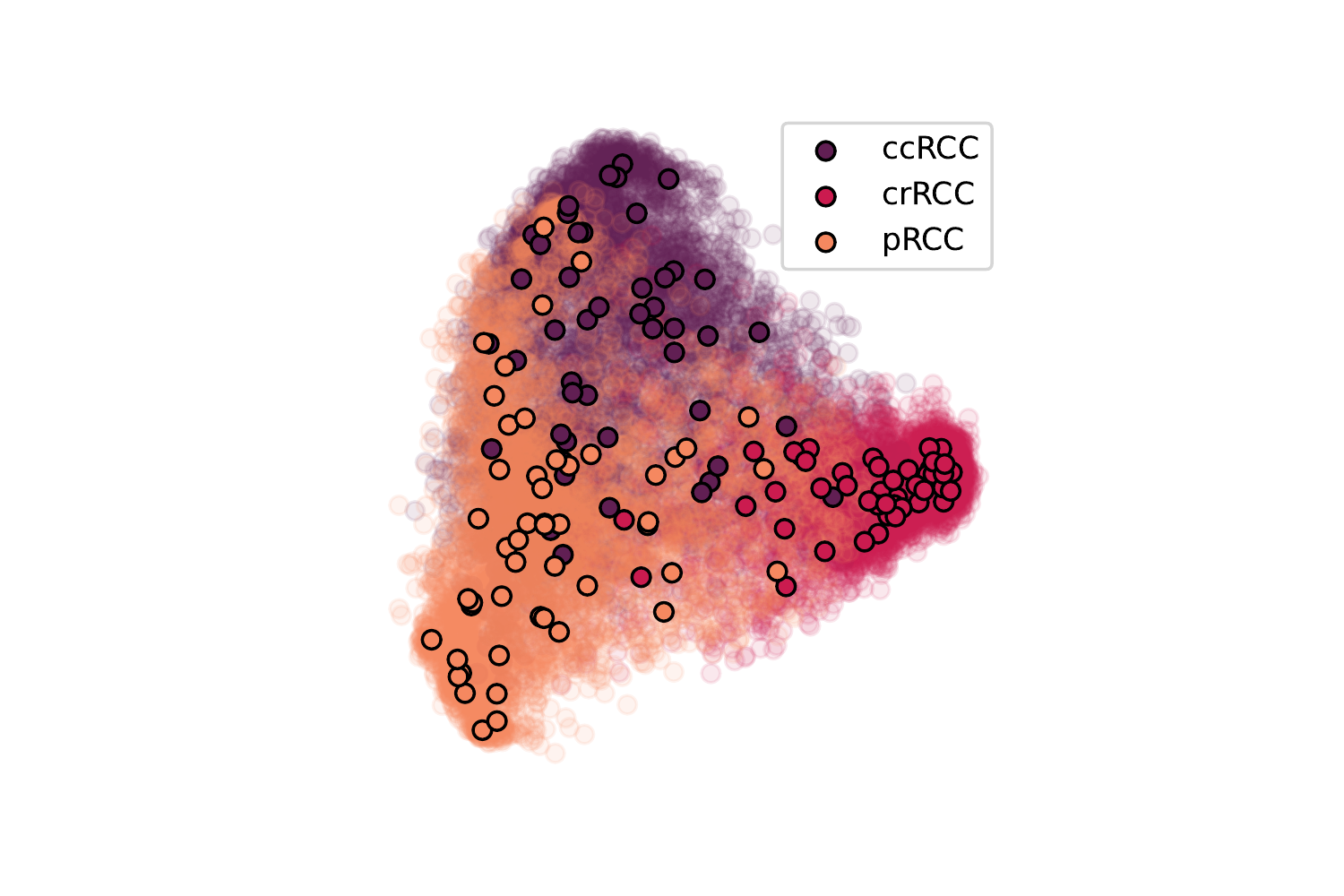}
    \caption{$\textrm{tRNAsformer}_{L=2}$}
    \end{subfigure}\\
    \begin{subfigure}{0.4\textwidth}
    \includegraphics[width=1\linewidth]{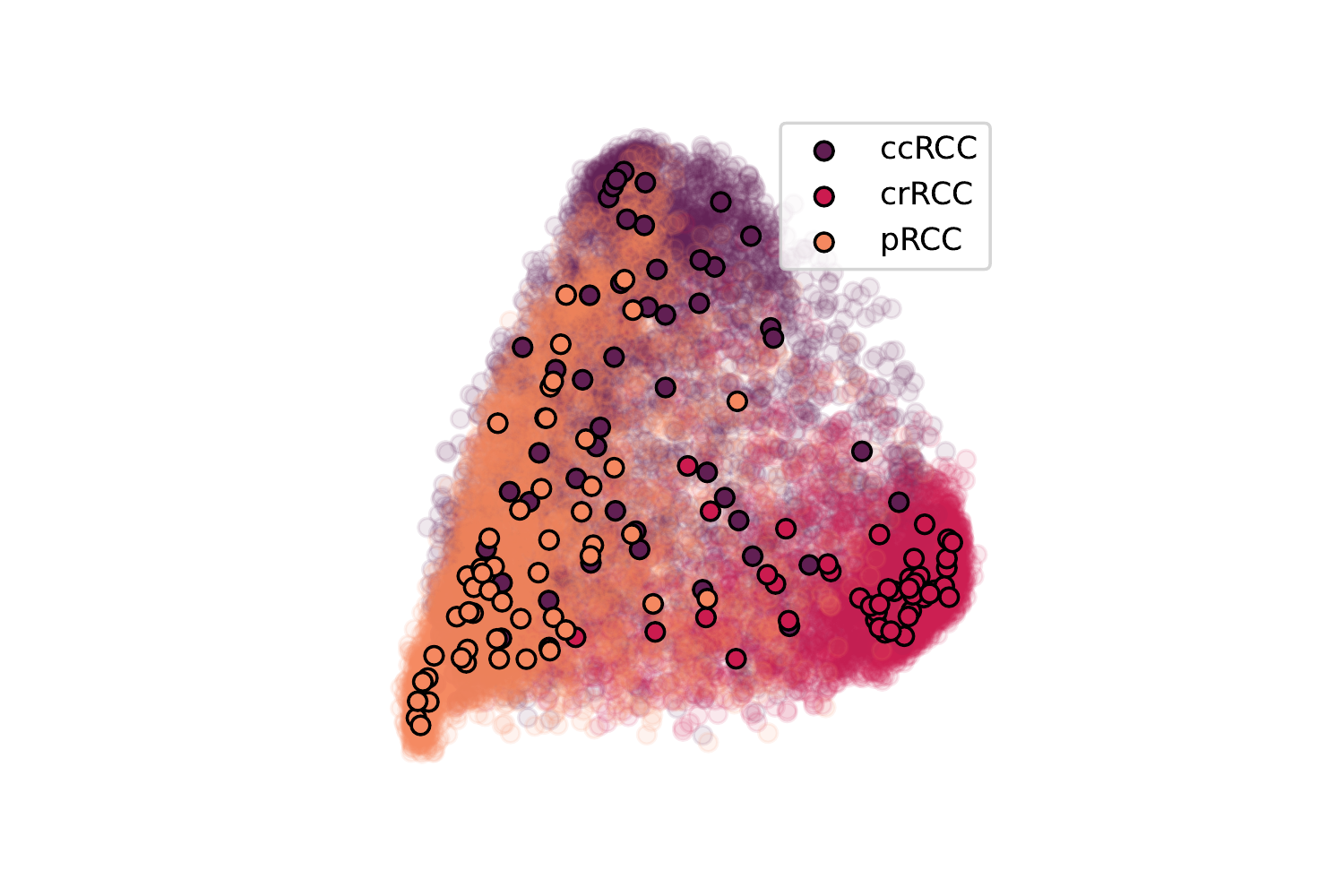}
    \caption{$\textrm{tRNAsformer}_{L=4}$}
    \end{subfigure}
    \begin{subfigure}{0.4\textwidth}
    \includegraphics[width=1\linewidth]{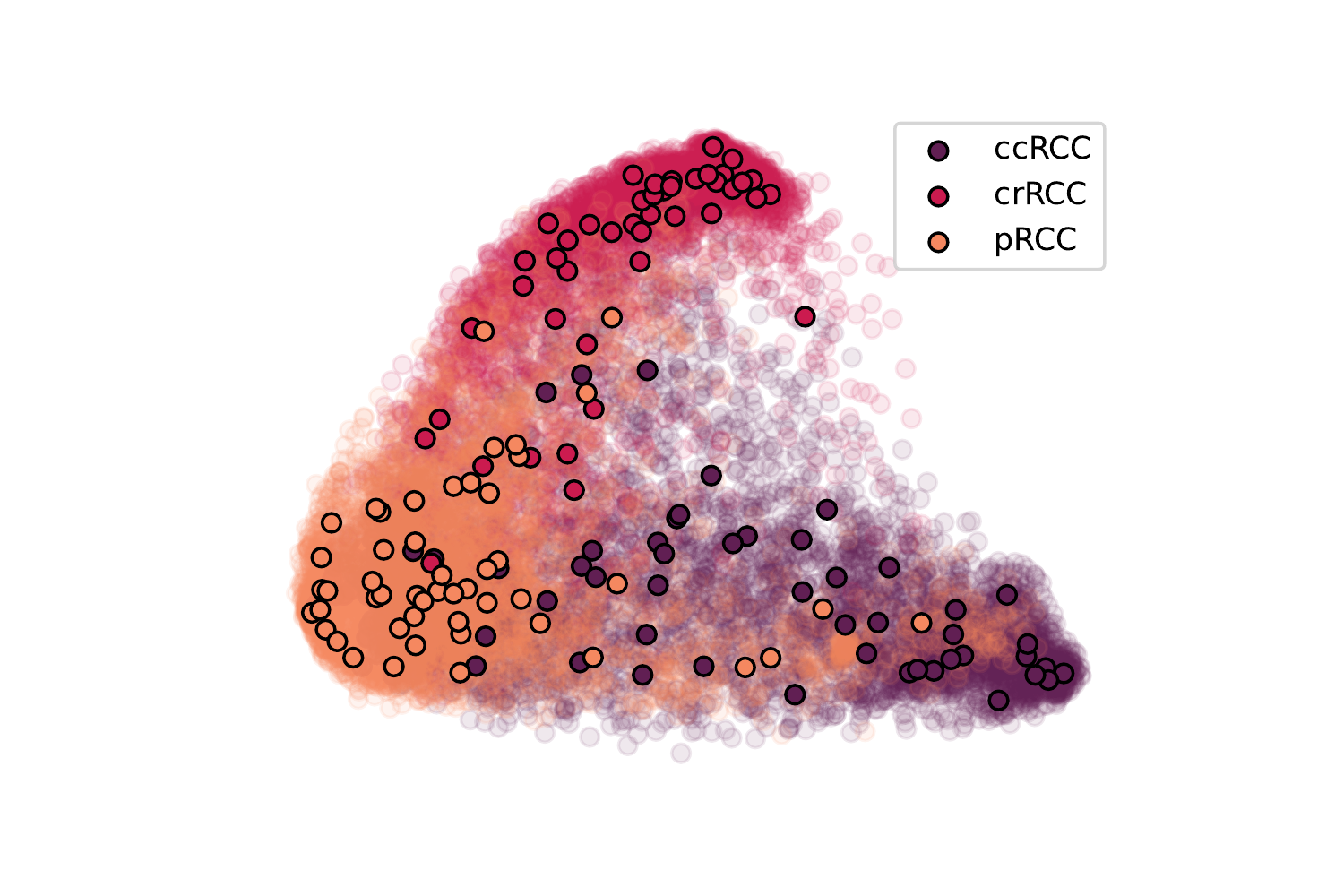}
    \caption{$\textrm{tRNAsformer}_{L=8}$}
    \end{subfigure}\\
    \begin{subfigure}{0.4\textwidth}
    \includegraphics[width=1\linewidth]{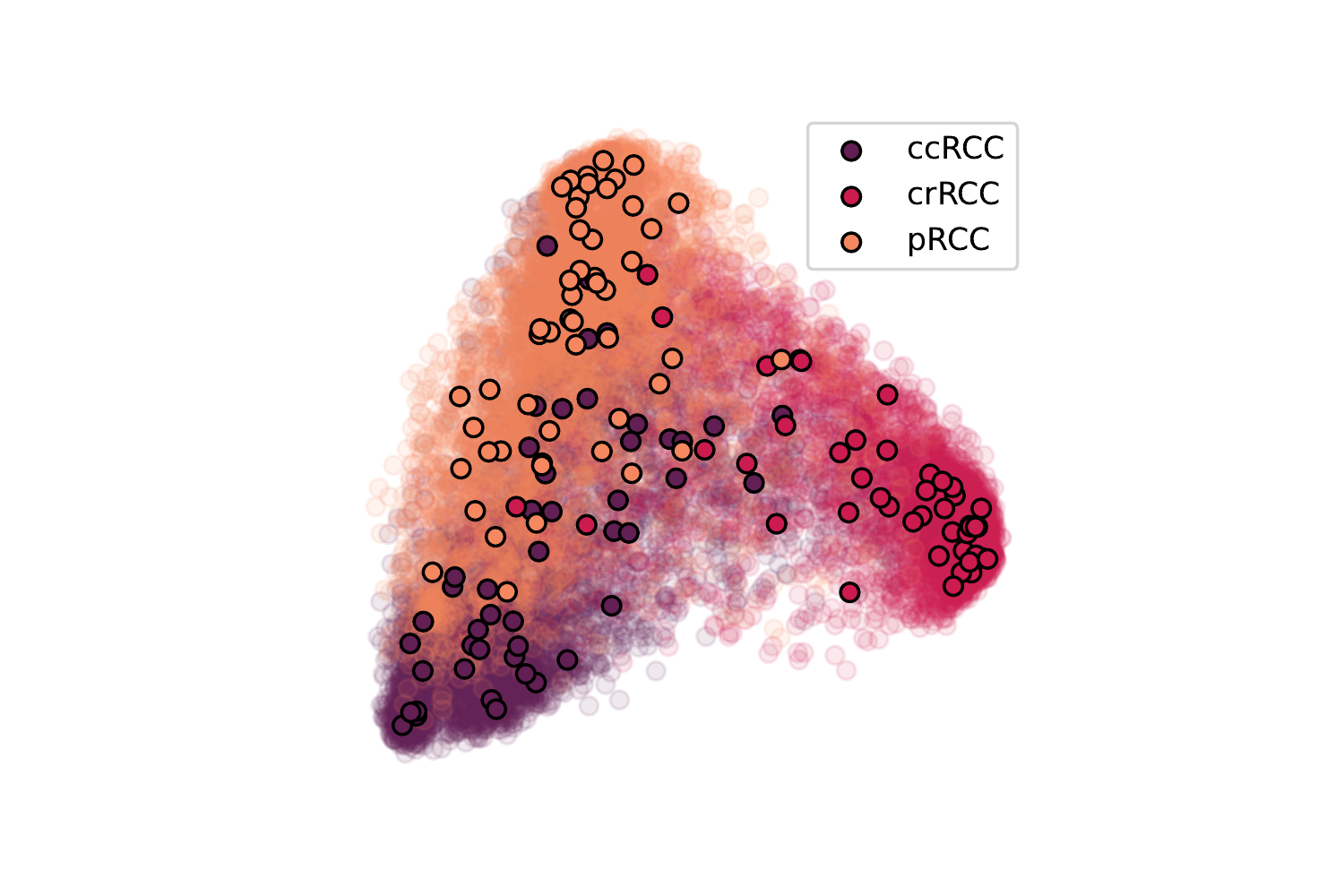}
    \caption{$\textrm{tRNAsformer}_{L=12}$}
    \end{subfigure}
    \caption{The two-dimensional PCA projection of the external dataset WSI features. (a)-(f) are for $\textrm{tRNAsformer}_{L}$, $L=(1,2,4,8,12)$, respectively. Each external test WSI is represented by 100 bags of features. All bags of features associated with the test set are shown with transparent circles. The average of PCA projection of each WSI (average of 100 bags associated with each WSI) is shown in bold circles with black edges.}
    \label{fig:rna-pca-external}
\end{figure}

\end{document}